\documentclass{ieeeaccess}
\usepackage{cite}
\usepackage{amsmath,amssymb,amsfonts}
\usepackage{algorithmic}
\usepackage{graphicx}
\usepackage{epstopdf}
\usepackage{textcomp}
\usepackage{amsmath, amsthm, amssymb, amsfonts, mathtools}
\usepackage{cases, multirow}
\usepackage{cite}
\usepackage{courier}
\usepackage{subfigure}
\usepackage{float}
\usepackage{color}
\usepackage{multicol}
\usepackage{dblfloatfix}
\usepackage{cuted, nccmath}
\usepackage{lipsum}
\usepackage{mathtools}
\usepackage{threeparttable}
\usepackage{caption}
\usepackage{esint}
\usepackage{mathbbol}

\long\def\comment#1{}

\def\figref#1{Fig.~\ref{#1}}
\def\be{\begin{equation} }
\def\ee{\end{equation}}

\DeclareCaptionFont{ieeeblue}{\color{accessblue}}
\DeclareCaptionLabelFormat{myformat}{\figcapfont{\textbf{#1}\textbf{#2}}}
\usepackage[font={small}]{caption}
\def\BibTeX{{\rm B\kern-.05em{\sc i\kern-.025em b}\kern-.08em
    T\kern-.1667em\lower.7ex\hbox{E}\kern-.125emX}}
\begin{document}
\history{Date of publication xxxx 00, 0000, date of current version xxxx 00, 0000.}
\doi{10.1109/ACCESS.2017.DOI}

\title{AI-Based Channel Prediction in D2D Links: An Empirical Validation}
\author{\uppercase{Nidhi Simmons}\authorrefmark{1}, \IEEEmembership{Member, IEEE},
\uppercase{Samuel B. Ferreira Gomes}\authorrefmark{2},
\uppercase{Michel Daoud Yacoub}\authorrefmark{2},
\IEEEmembership{Member, IEEE},
\uppercase{Osvaldo~Simeone}\authorrefmark{3}, \IEEEmembership{Fellow, IEEE},
\uppercase{Simon~L.~Cotton}\authorrefmark{1}, \IEEEmembership{Senior Member, IEEE}, and
\uppercase{David E. Simmons}\authorrefmark{4}
}
\address[1]{Institute of Electronics, Communications and Information Technology, Queen's University Belfast, BT3 9DT, UK (e-mail: \{nidhi.simmons, simon.cotton\}@qub.ac.uk)}
\address[2]{Wireless Technology Laboratory, School of Electrical and Computer Engineering, University of Campinas, Campinas 13083-970, Brazil (e-mail: samuelbf@decom.fee.unicamp.br and mdyacoub@unicamp.br)}
\address[3]{King's Communications, Learning and Information Processing (KCLIP), Centre for Telecommunications Research, Department of Informatics, King's College, London, WC2R 2LS, UK (e-mail: osvaldo.simeone@kcl.ac.uk)} 
\address[4]{Senior software engineer working in the area of machine learning in Belfast, UK, (e-mail: dr.desimmons@gmail.com)}
\tfootnote{This work was supported by the Royal Academy of Engineering under Grant Reference RF\textbackslash201920\textbackslash 19\textbackslash 191.}

\markboth
{N. Simmons \headeretal: AI-Based Channel Prediction in D2D Links: An Empirical Validation}
{N. Simmons \headeretal: AI-Based Channel Prediction in D2D Links: An Empirical Validation}

\corresp{Corresponding author: Nidhi Simmons (e-mail: nidhi.simmons@qub.ac.uk).}

\begin{abstract}
Device-to-Device (D2D) communication propelled by artificial intelligence (AI) will be an allied technology that will improve system performance and support new services in advanced wireless networks (5G, 6G and beyond).
In this paper, AI-based deep learning techniques are applied to D2D links operating at 5.8~GHz with the aim at providing potential answers to the following questions concerning the prediction of the received signal strength variations: \textit{i) how effective is the prediction as a function of the coherence time of the channel?} and \textit{ii) what is the minimum number of input samples required for a target prediction performance?} To this end, a variety of measurement environments and scenarios are considered, including an indoor open-office area, an outdoor open-space, line of sight (LOS), non-LOS (NLOS), and mobile scenarios. Four deep learning models are explored, namely long short-term memory networks (LSTMs), gated recurrent units (GRUs), convolutional neural networks (CNNs), and dense or feedforward networks (FFNs). Linear regression is used as a baseline model. It is observed that GRUs and LSTMs present equivalent performance, and both are superior when compared to CNNs, FFNs and linear regression. This indicates that GRUs and LSTMs are able to better account for temporal dependencies in the D2D data sets. 
We also provide recommendations on the minimum input lengths that yield the required performance given the channel coherence time. For instance, to predict 17~and~23~ms into the future, in indoor and outdoor LOS environments, respectively, an input length of 25~ms is recommended. This indicates that the bulk of the learning is done within the coherence time of the channel, and that large input lengths may not always be beneficial.

\end{abstract}


\titlepgskip=-15pt

\maketitle

\section{Introduction}
\label{sec:introduction}
Advanced wireless networks\textemdash{5G, 6G, and beyond}\textemdash aim to provide a multitude of new services, many of them supporting low-latency communications. One way to deliver low-latency communications is to enable wireless terminals to communicate with each other through direct, usually short device-to-device (D2D) links. These links can also be conveniently used to provide for additional features such as increase reliability and/or extension of radio coverage~\cite{dholer}.

The characteristics of signal propagation in D2D links can often be very different to those encountered in traditional wireless communications. In the latter, the base station is fixed with antennas elevated above rooftops, and the link is therefore, usually, free of local scatterings. D2D communications occur in an infrastructure-less network, with fixed or mobile terminals at low elevations, immersed in rich scattering environments. Furthermore, mobile terminals may be in close proximity to the human body, whose motion is bound to cause stochastic shadowing on the radio link~\cite{686781, 5782245}. Shadowing and scattering may be caused by a number of different radio obstructions present in the local environment, such as vehicles and buildings (outdoor), internal walls and furniture (indoor), and pedestrians (indoor and outdoor)~\cite{article}. 

This paper is concerned with the problem of predicting radio channel conditions encountered by D2D links. Given the difficulty in modelling D2D links, we explore the use of artificial intelligence (AI) tools and rely on extensive experiments.

\subsection{Related Work}

Several works have previously addressed the wireless channel prediction problem using AI techniques~\cite{8057090, 8653275, 8761432 , 8884240}. In~\cite{8057090}, a hybrid deep learning model for spatiotemporal prediction in cellular networks was presented. It was shown that the proposed model, which included an autoencoder-based deep network for spatial modeling and long short-term memory network (LSTM) for temporal modeling, significantly improved prediction accuracy when compared to two commonly used baseline methods, namely autoregressive integrated moving average, and support vector regression. Channel prediction using LSTMs and autoregressive methods was also applied to vehicular measurements in~\cite{8653275}. Unlike in~\cite{8057090}, where it was observed that ‘learning more’ (i.e., increasing the number of stacked layers and hidden units in each layer) helped improve prediction performance, it was found in~\cite{8653275} that LSTMs with just a small number of hidden units performed better when compared to increasing the number of hidden units in each layer. Reference~\cite{8653275} only compared a single deep learning model with a single baseline model. 

An initial study on channel prediction in body area networks (BANs) was carried out in~\cite{8761432}. It was shown that an LSTM based framework performed better on BAN measurements when compared with existing approaches such as moving average and adaptive prediction. However, these studies did not include comprehensive experiments using real data from a mobile device. Wireless channel quality prediction was also studied in~\cite{8884240} where an encoder-decoder based sequence-to-sequence deep learning model was used, and its performance was compared with linear regression for multiple networks and communication standards. It was observed that sequence lengths of size 20 captured most of the useful information in the data, and sequences of greater lengths did not improve prediction performance. 

Concerning D2D communications, reference~\cite{8891760} focused on deep learning approaches for content caching in cache-enabled D2D networks. Two recurrent neural network approaches, namely echo state networks and LSTMs, were employed to predict users’ mobility and content popularity, so as to determine which content to cache and where to cache. However, these results were not tested on real-world channel measurements. In~\cite{9142437}, a deep learning approach was proposed to predict D2D channel gains from independent cellular channel gains in order to solve various problems related to radio resource management. All predictions were based on the assumption of the channel being Gaussian.

\subsection{Main Contributions}

This paper explores AI-based deep learning techniques on D2D links with the aim at providing potential answers to the following questions concerning the prediction of the received signal strength variations:
\begin{enumerate}
    \item How effective is the prediction as a function of the coherence time of the channel?
\item What is the minimum number of input samples required for a target prediction performance? 
\end{enumerate}
To address these questions, four deep learning models are explored. These include dense or feedforward networks (FFNs), convolutional neural networks (CNNs), gated recurrent units (GRUs), and LSTMs. It is worth mentioning that CNNs are commonly applied to analyse image data, but they also find application in predicting time series data~\cite{goodfellow2016deep}. Furthermore, recurrent neural networks (RNNs) are powerful in discovering the dependency in sequential data. Specifically, GRUs and LSTMs work well on sequential data with long-term dependencies~\cite{lin2017hybrid, chung2014empirical, LongShort} due to their internal memory mechanisms. 

Accordingly, unlike prior work, we focus on understanding empirically the relationship between channel coherence time and number of samples used in the prediction models, as well as the minimum input length required to achieve a target prediction performance for a given coherence time.  To this end, we compare and validate the prediction performances of the deep learning models on real-world D2D field measurements conducted for a variety of environments and scenarios at 5.8~GHz. These include an indoor open office  area environment, an outdoor open space environment, LOS, non-LOS (NLOS) and mobile scenarios. 

The remainder of this paper is organized as follows. Section~II describes the D2D channel measurements conducted at 5.8 {GH}z. Section~III discusses the the data preprocessing steps applied. Section~IV discusses the deep learning models used for prediction and their implementation details. Section~V presents the experimental results. Lastly, Section~VI finishes the paper with some concluding remarks.

\begin{figure}[t]
\centering
\includegraphics[width=3.5in]{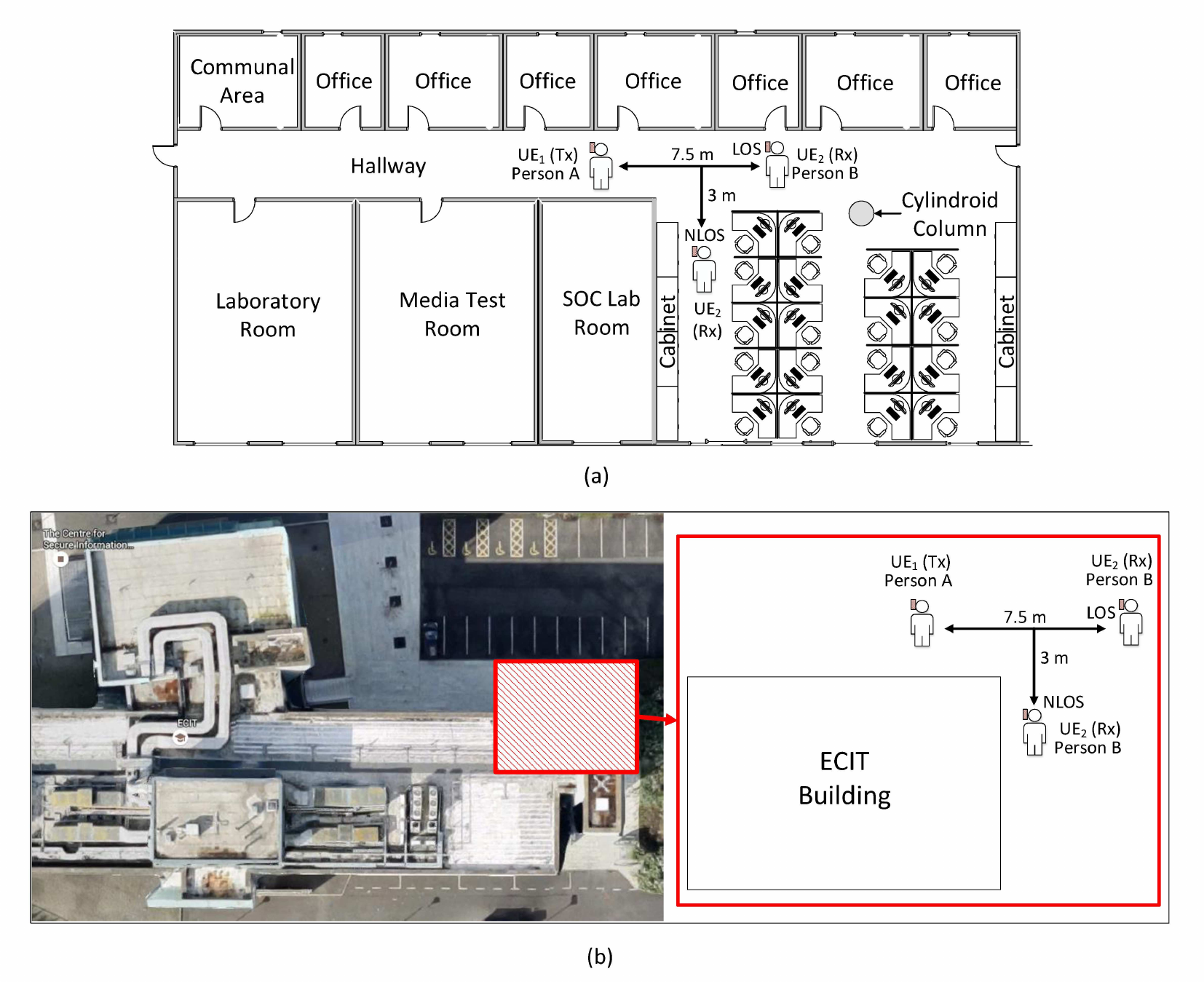}
\caption{D2D measurements in (a) an indoor open office environment and (b)~an outdoor open space environment showing different locations of person~B for the LOS and NLOS cases.}
\label{fig:img1}
\vspace{-0.2cm}
\end{figure}
\begin{figure*}[b]
\centering
\subfigure{{\includegraphics[width=8cm]{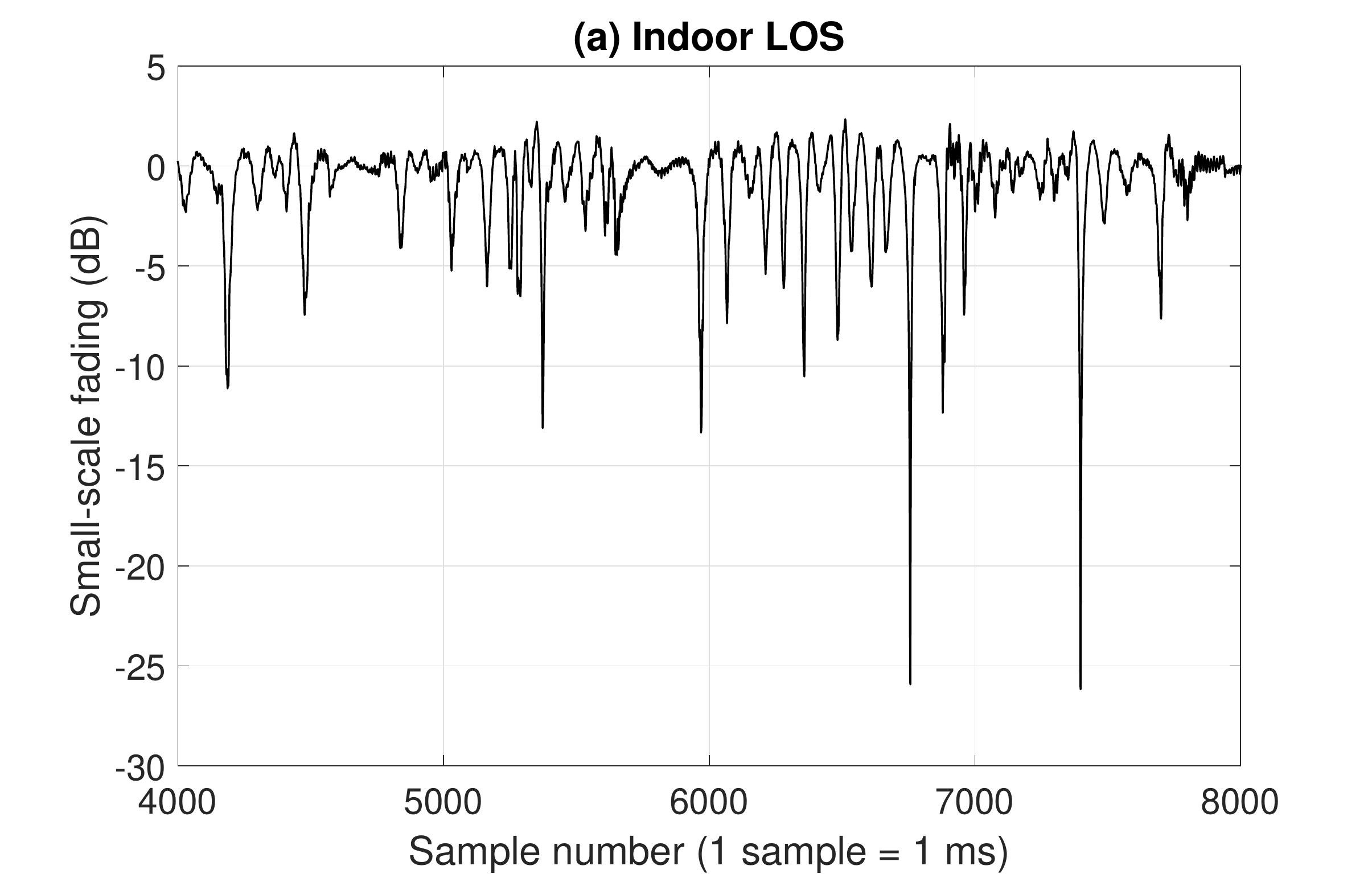} }}
\quad
\subfigure{{\includegraphics[width=8cm]{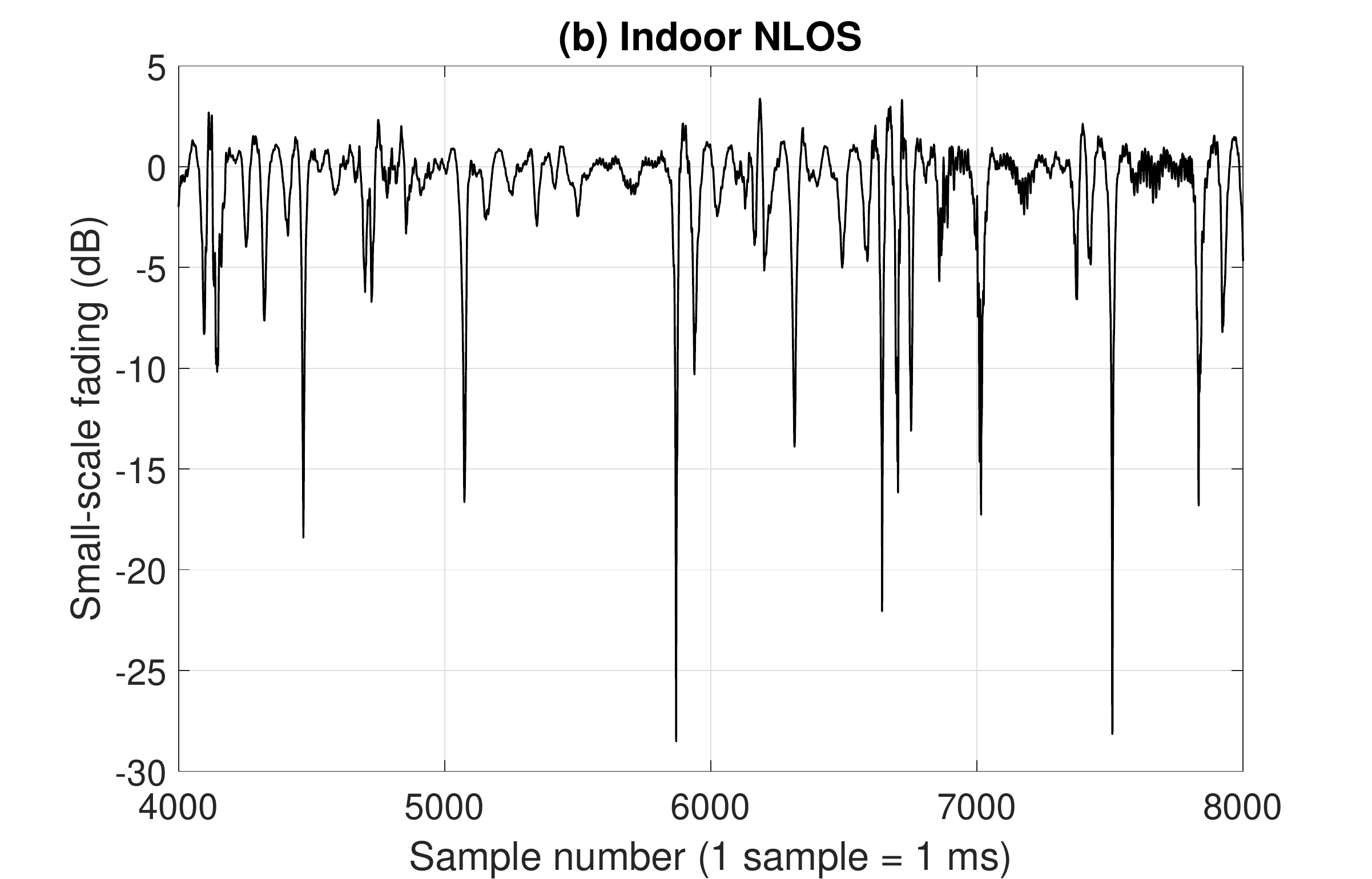} }}

\subfigure{{\includegraphics[width=8.05cm]{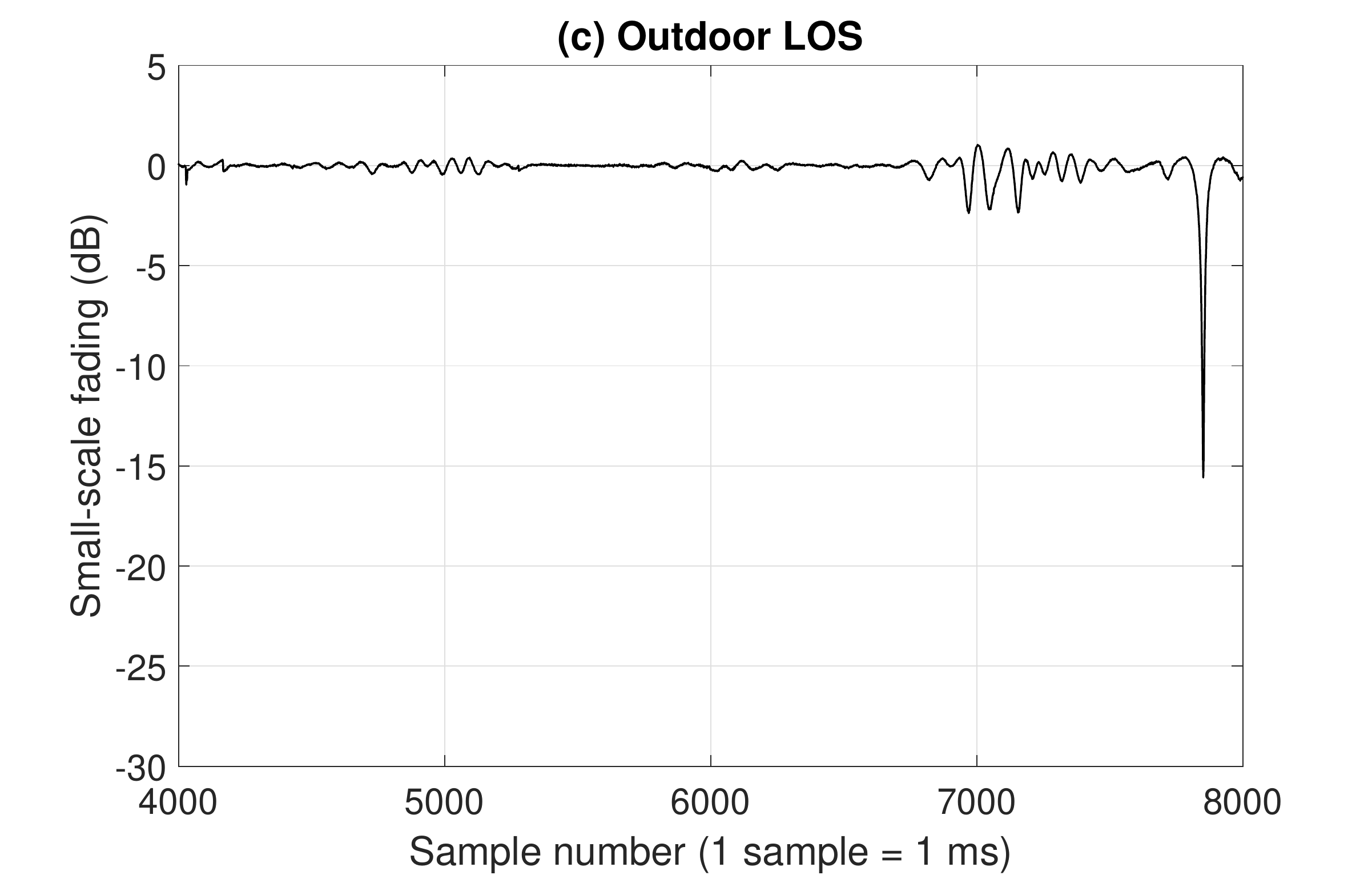}}}
\quad
\subfigure{{\includegraphics[width=8cm]{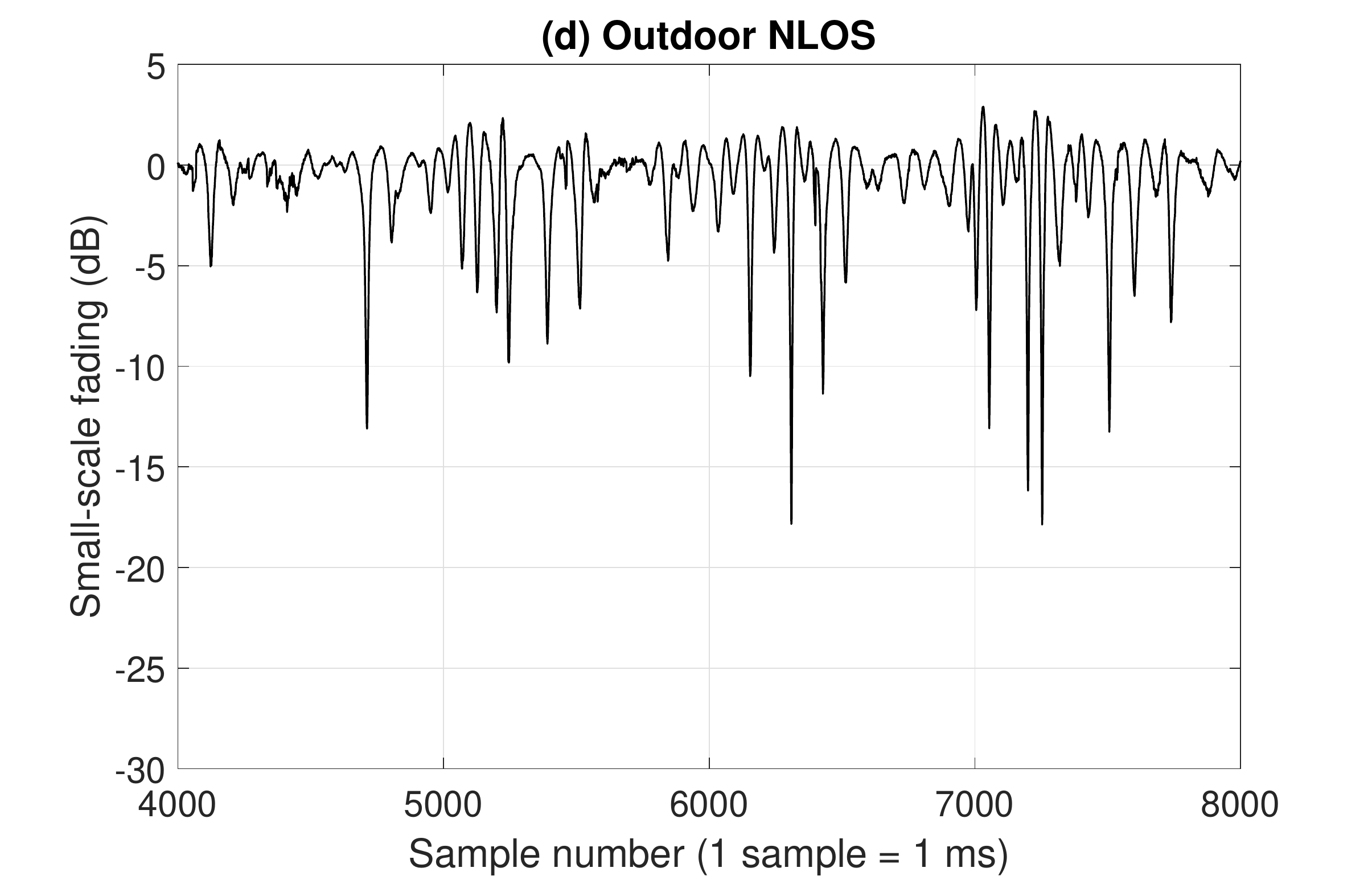}}}
\caption{Small-scale fading variations in (a) Indoor LOS environment (b) Indoor NLOS environment (c)~Outdoor LOS environment and (d)~Outdoor NLOS environment.}
\label{fig:img2}
\vspace{-0.2cm}
\end{figure*}

\section{D2D Channel Measurements}
The wireless channel measurement system used in this study was based on the ML5805 transceivers, manufactured by RFMD~(Qorvo)\footnote{https://ir.qorvo.com/news-releases/news-release-details/rfmdr-releases-industrys-first-58-ghz-ism-band-transceiver}. The transceiver boards were interfaced with a PIC32MX which acted as a baseband controller and allowed the analog received signal strength (RSS) to be sampled with a 10-bit quantization depth. A D2D link was formed between two persons, namely person~A (an adult female of height 1.65~m and weight 53~kg) and person~B (an adult male of height 1.83~m and weight 73~kg). The user equipment (UE) positioned on person~A acted as the transmitter and was configured to output a continuous wave signal with a power level of +17.6 dBm at 5.8~{GH}z. The UE positioned on person~B acted as the receiver and sampled the channel at a rate of 10~k{H}z. 

The RSS data was downsampled by averaging 10 consecutive samples to improve the signal to noise ratio (SNR) performance, thus giving an effective sampling rate of 1~k{H}z after downsampling.
Furthermore, the antennas used by the transmitter and receivers were +2.3~d{B}i sleeve dipole antennas (Mobile Mark model PSKN3-24/55S). The antennas were housed in a compact acrylonitrile butadiene styrene (ABS) enclosure (107 × 55 × 20 mm). This setup was representative of the form factor of a smart phone which allowed the user to hold the device as they normally would to make a voice call. Each antenna was securely fixed to the inside of the enclosure using a small strip of Velcro®. 

The D2D channel measurements were obtained within an indoor open office area and an outdoor open space environment, as shown in~\figref{fig:img1}. The indoor open office was located on the first floor of the Institute of Electronics Communications and Information Technology (ECIT) building at Queen’s University Belfast in the United Kingdom. The building mainly consists of metal studded dry walls with metal tiled floors covered with polypropylene-fiber, rubber backed carpet tiles, a metal ceiling with mineral fiber tiles and recessed louvered luminaries suspended 2.7~m above floor level. The office contained a number of chairs, metal storage spaces, doors and desks constructed from medium density fibreboard. These desks were vertically separated by soft wooden partitions. During the measurements, the office area was unoccupied in order to facilitate pedestrian free D2D channel measurements. The outdoor D2D measurements were conducted in an outdoor car parking area adjacent to the ECIT building.

As shown in~\figref{fig:img1}, during the D2D measurements~\cite{7467556}, person~A and~B held their UE at their left-ears to imitate making a voice call. For the LOS D2D measurements, person~B was positioned directly in front of person~A whilst for the NLOS D2D measurements, person~B was positioned around an adjacent corner. It is worth noting that both test subjects were initially stationary after which they were instructed to walk around randomly within a circle of radius 0.5~m from their starting points. For the LOS D2D measurements in both environments, while there may have been a direct LOS between the two person’s bodies during the trials, in actual fact, the link between the hypothetical UEs would have been subject to quasi-LOS conditions due to the random movements undertaken. For the NLOS case, person B was always positioned around an adjacent corner to ensure that the NLOS conditions (i.e. no direct signal path between persons~A and~B) were maintained irrespective of the random movements.

\section{Data Preprocessing}

Once the RSS measurements were obtained, the small-scale fading data was extracted for analysis. Specifically, the large-scale fading component was removed by applying a low-pass filter to the raw RSS data in linear scale. To determine the window size for extraction of the local mean signal, the raw data was visually inspected and overlaid with the local mean signal for differing window sizes. A smoothing window of 50~samples was then used.
\figref{fig:img2} shows the small-scale fading variations for a 4~s window observed in the indoor LOS, indoor NLOS, outdoor LOS and outdoor NLOS environments, respectively. The overall measurement data set consisted of approximately 62300 samples, equivalently 62.3~s length in time. This included the measurement data for all scenarios.

Data scaling was then applied to transfer the data into ranges and forms that are appropriate for modeling. It is well known that models trained on scaled data perform significantly better when compared to models trained on unscaled data~\cite{guyon2003introduction}. As well as this, the gradient descent converges much faster with scaled data than without it~\cite{aksoy2001feature}. In this paper, the data sets were scaled using min-max normalisation (which performs a linear transformation on the original  data)~\cite{minmax} before being input to the model for training.
Let $x_{min}$, and $x_{max}$ be the minimum and  maximum values for attribute $X$. Min–max normalization maps a value $v$ of $X$ to $\acute{v}$ in the range [${\rm{new}}~x_{min}$, ${\rm{new}}~x_{max}$] using~\eqref{eq000}, as follows:

 \begin{equation}
 \acute{v}\! = \frac{v - x_{min}}{x_{max}-x_{min}} \left({\rm{new}}~{x_{max}} - {\rm{new}}~{x_{min}}\right) + {\rm{new}}~{x_{min}}.  
 \label{eq000}
 \end{equation}
Note that the normalization output was customised to be in the range [-1, 1] by rewriting~\eqref{eq000}, as follows 
 \begin{equation}
 \acute{v} = 2\frac{v - x_{min}}{x_{max}-x_{min}} -1.  
 \label{eq001}
 \end{equation}

\section{Methodology}
This work focuses on a univariate time series forecasting problem. Here, data sets comprised of only a single variable are observed at each time step, and a model is used to exploit the values seen at prior time steps to predict the subsequent time step values. A \textit{sliding window}\footnote{Is a statistical method in which a window of specified length moves over the data, sample by sample, and the statistic is computed over the data in the window.} approach is adopted  to restructure the time series data as a supervised learning problem.\footnote{Supervised learning is the most popular way of framing problems for machine learning as a collection of observations with inputs and outputs.} Thus, the models here make a set of predictions based on a window of consecutive samples from the data sets.

This section first discusses how the prediction problem in this paper is framed in a supervised learning manner through data windowing. Then, the baseline and deep learning models used here are explained. Following this, their implementation details are presented.

\subsection{Data Windowing}
\begin{figure*}[h]
\centering
\includegraphics[width=7in]{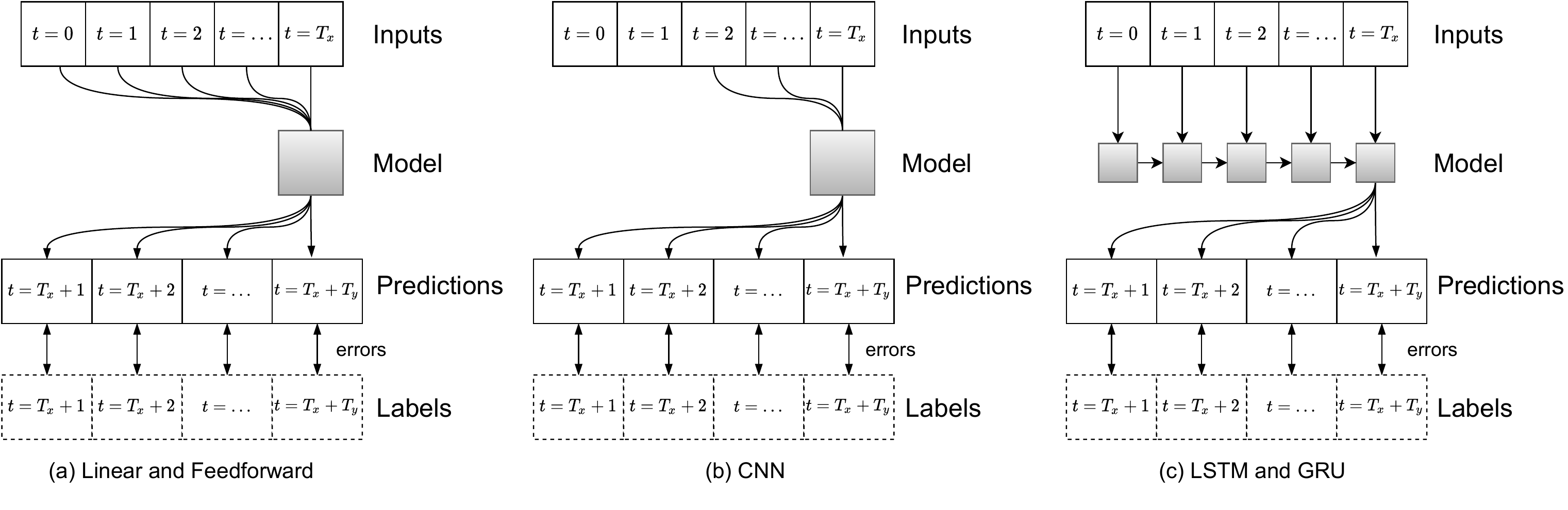}
\caption{Input-output relationship in (a) Linear and Feedforward models (b) CNN model with kernel size of 3 and (c) LSTM and GRU models.}
\label{fig:i/o_models}
\end{figure*}

Data windowing of the models is represented in ~\figref{fig:i/o_models}. The input size, also called as the input width, is the number of time steps considered by the window as an input, and is denoted by $T_{x}$. The number of output steps to be predicted, also called as the horizon, is represented as $T_{y}$.

Linear and feedforward flatten the input data as a vector to convey the previous time steps with size $T_{x}$. The main drawback of this approach is that the resulting model can only be executed on input windows of exactly the same shape. The CNN model also takes multiple time steps as input to produce one-shot $T_{y}$ out steps predictions. However, different than feedforward networks, CNNs can be run on inputs of any length, and the predictions are based on a fixed-width history controlled by their kernel sizes. This might result in better performance  
\noindent
since it can see how things are changing over time.
Recurrent neural networks such as LSTMs and GRUs are intrinsically well-suited to process sequential data. This is done by maintaining an internal state from time step to time step. At each time step an input size of $T_{x}$ is fed into the model producing $T_{y}$ output steps as predictions. For the next time step, the data window is shifted by $T_{y}$ samples.

\subsection{Linear Regression Baseline Model}
This model assumes that the relationship between the independent variables (or features) $\boldsymbol{x}$ and the dependent variable $y$ is linear i.e., $y$ can be expressed as a weighted sum of the elements in $\boldsymbol{x}$, given some noise on the observations. It should be noted that the baseline model here refers to a univariate linear regression model. 

Assuming that the inputs consist of $T_{x}$ features, the prediction ${\hat{y}}$ is expressed as
\begin{equation}
    {\hat{y}} = {w_{1}}{x_{1}}+\dots+{w_{T_{x}}}{x_{T_{x}}}+b
    \label{linear_eq001a}
\end{equation}
where $w_{1}, \dots, w_{T_{x}}$ are called weights, and $b$ is called a bias (also called an offset or intercept). The weights determine the influence of each feature on the prediction, and the bias indicates the value that the prediction should take when all of the features take value 0. Models whose output prediction is determined by the affine transformation of input features are linear models, where the affine transformation is specified by the chosen weights and bias~\cite{zhang2020dive}. Now collecting all features into a vector $\boldsymbol{x} \in {\mathbb{R}}^{T_{x}}$ and all weights into a vector $\boldsymbol{w} \in {\mathbb{R}}^{T_{x}}$, the model in~\eqref{linear_eq001a} can be expressed as~\cite[eq. 3.1.3]{zhang2020dive},
\begin{equation}
    {\hat{y}} = {\boldsymbol{w}^\top}{\boldsymbol{x}} + b.
    \label{linear_eq001b}
\end{equation}
Here, the vector $\boldsymbol{x}$ corresponds to features of a single data example and $(\cdot)^\top$ is the vector transpose.  For a collection of features fed into the model in a batch size\footnote{It is the number of samples processed before the model is updated.} of $N$, $\boldsymbol{X} \in {\mathbb{R}}^{N \times T_{x}}$ and $\boldsymbol{w}\in\mathbb{R}^{T_{x}\times T_{y}}$, the predictions ${\boldsymbol{\hat{y}}} \in {\mathbb{R}}^{N \times T_{y}}$, can be expressed via the matrix-vector product~\cite[eq. 3.1.4]{zhang2020dive}
\begin{equation}
    {\boldsymbol{\hat{{y}}}} = {\boldsymbol{{X}}}{\boldsymbol{{w}}} + \boldsymbol{b}.
    \label{linear_eq001c}
\end{equation}
with $\boldsymbol{b}\in\mathbb{R}^{1\times T_{y}}$.

\begin{figure}[h]
\centering
\includegraphics[width=2.4in]{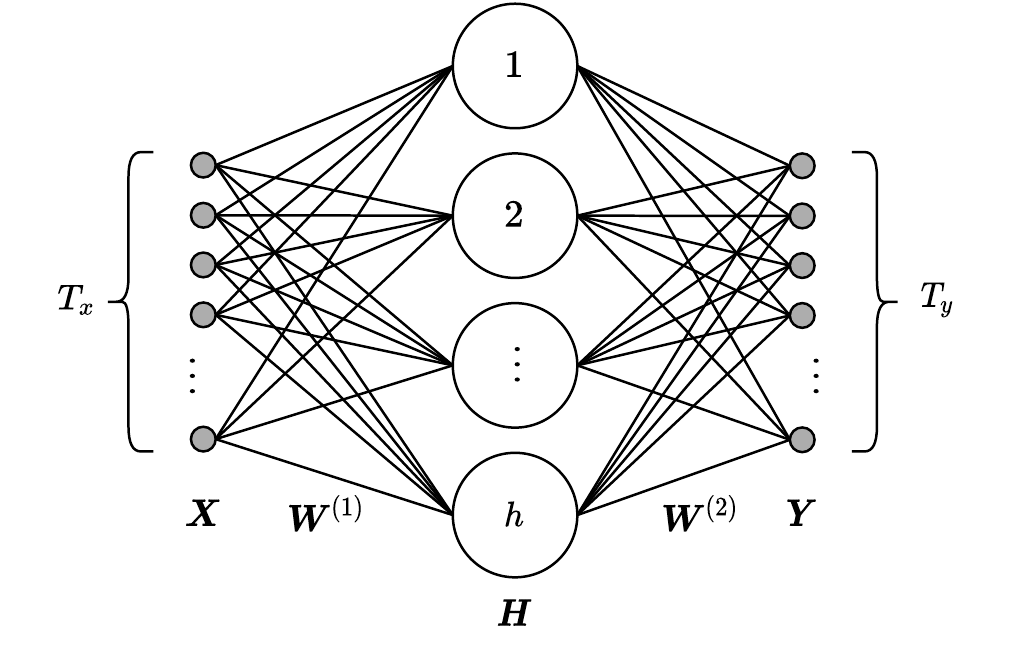}
\caption{Dense/feed-forward network with one hidden layer with $h$ hidden units.}
\vspace{-0.1cm}
\label{fig:dense}
\end{figure}
\subsection{Feedforward Network}
These networks are also known as dense networks, and are capable of handling a more general class of functions by incorporating one or more hidden layers. These layers create non-linear representations of the data and are able to capture complex interactions among the input. The final (output) layer is usually a linear predictor. 

The network layers are connected in a fully connected manner, meaning that every input influences every neuron in the hidden layer, and each of these influence every neuron in the output layer. A dense network with one hidden layer is illustrated in~\figref{fig:dense}. The inputs $\boldsymbol{X}\in\mathbb{R}^{N\times T_{x}}$ are being fed into the model in a batch size of $N$ training instances where each instance has $T_{x}$ inputs. Considering one hidden layer network whose hidden layer has $h$ hidden units, the hidden representation, $\boldsymbol{H}\in\mathbb{R}^{N\times h}$, and the network output, $\boldsymbol{Y}\in\mathbb{R}^{N\times T_{y}}$, are given as~\cite[eq. 4.1.3]{zhang2020dive}
\begin{equation}
    {\boldsymbol{H}}={g\left(\boldsymbol{X}\boldsymbol{W}^{(1)}+\boldsymbol{b}^{(1)}\right)} 
    \label{dense_eq001}
\end{equation}
and
\begin{equation}
    {\boldsymbol{Y}}={\boldsymbol{H}\boldsymbol{W}^{(2)}+\boldsymbol{b}^{(2)}} ,
    \label{dense_eq002}
\end{equation}
respectively. The weights and biases of the hidden layer are $\boldsymbol{W}^{(1)}\in\mathbb{R}^{T_{x}\times h}$ and $\boldsymbol{b}^{(1)}\in\mathbb{R}^{1\times h}$, respectively, whereas the weights and biases of the output layer are $\boldsymbol{W}^{(2)}\in\mathbb{R}^{h\times T_{y}}$ and $\boldsymbol{b}^{(2)}\in\mathbb{R}^{1\times T_{y}}$, respectively. Finally, the activation function $g(\cdot)$ is responsible for introducing non-linearity in the model. In this work, we adopt rectified linear units (ReLU), $g(x)=\max\{0,x\}$, as the hidden layer activations.

\subsection{Long Short-Term Memory Network}
The LSTM network is a type of RNN that is well known for its time series prediction capabilities. In a standard RNN, the nodes i.e., the building blocks of a neural network architecture are composed of basic activation functions such as tanh and sigmoid. As indicated in~\cite{goodfellow2016deep}, because RNN weights are learned by backpropagating errors through the network, the use of these activation functions can cause RNNs to suffer from the vanishing gradient problem that causes the gradient to have either infinitesimally low or high values. This affects a recurrent neural network’s ability to learn long-term dependencies~\cite{818041}. The LSTM network is able to partially overcome the vanishing gradient problem by creating paths through time that have derivatives that neither vanish nor explode~\cite{goodfellow2016deep} by incorporating the ability to forget.
\begin{figure}[t]
\centering
\includegraphics[width=3.6in]{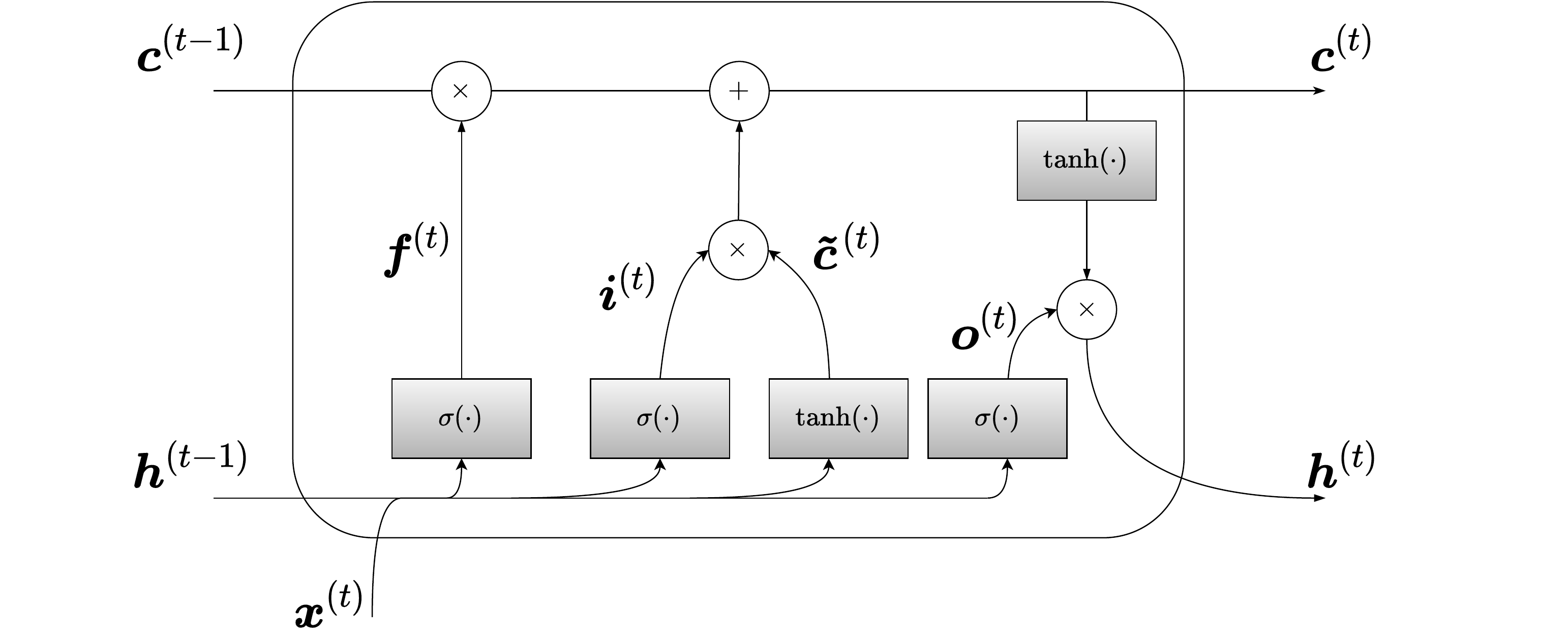}
\caption{Block diagram of the LSTM cell.}
\label{fig:lstm}
\vspace{-0.2cm}
\end{figure}

As explained in~\cite{goodfellow2016deep}, LSTM recurrent networks have LSTM cells, which includes a memory cell (or cell for short), designed to record additional information (which allows it to handle long-term dependencies). Each cell has the same inputs and outputs as an ordinary recurrent network, and also has more parameters and a system of three gating units that controls the flow of information, namely the output gate, input gate, and forget gate. These gates were specifically designed inspired by logic gates of a computer. The output gate reads out the entries from the cell. The input gate decides when to read data into the cell. Lastly, the forget gate represents a mechanism for resetting the cell's content. The main motivation of this gating design is to be able to decide when to remember and when to ignore inputs in the hidden state.~\figref{fig:lstm} shows the block diagram of a single LSTM cell which has an internal recurrence (a self-loop), in addition to the outer recurrence of the RNN. The most important component is the memory cell state unit ${\boldsymbol{c}^{(t)}\in\mathbb{R}^{N\times h}}$ that captures the internal state of the LSTM cell and has a linear self-loop given by~\cite[eq. 9.2.3]{zhang2020dive}

\begin{equation}
    {\boldsymbol{c}^{(t)}} = {\boldsymbol{f}^{(t)}}\odot{\boldsymbol{c}^{(t-1)}}+ {\boldsymbol{i}^{(t)}}\odot{\boldsymbol{\tilde{c}}^{(t)}},
    \label{eq002}
\end{equation}
where $\odot$ is the Hadamard (elementwise) product operator.

The memory cell is updated by partially forgetting the existing memory and adding a new memory content. This candidate memory cell ${\boldsymbol{\tilde{c}}^{(t)}\in\mathbb{R}^{N\times h}}$ represents the degree to which the new memory content is added to the memory cell and is modulated by the input gate ${\boldsymbol{i}^{(t)}\in\mathbb{R}^{N\times h}}$. The new memory content is given as~\cite[eq. 9.2.2]{zhang2020dive}

\begin{equation}
    {\boldsymbol{\tilde{c}}^{(t)}} = {\rm{tanh}}
\!\left(\!{\boldsymbol{x}^{(t)}}{\boldsymbol{W}}_{xc}+\! {\boldsymbol{h}^{(t-1)}}{\boldsymbol{W}_{hc}} \!+\! {\boldsymbol{b}_{c}}\!\right),
    \label{eq002_2}
\end{equation}
where ${\boldsymbol{W}_{xc}}\in\mathbb{R}^{T_{x}\times h}$ and ${\boldsymbol{W}_{hc}}\in\mathbb{R}^{T_{x}\times h}$ are input weight parameters and recurrent weights with respect to the cell gate, respectively, and ${\boldsymbol{b}_{c}}\in\mathbb{R}^{1\times h}$ is a bias parameter. The batch size is denoted by $N$, $T_{x}$ is the number of inputs, $\boldsymbol{x^{(t)}}\in\mathbb{R}^{N\times T_{x}}$ is the current input vector and $\boldsymbol{h^{(t)}}\in\mathbb{R}^{N\times h}$ is the current hidden layer vector with $h$ hidden units containing the outputs of all the LSTM cells. The self-loop weight is controlled by a forget gate unit ${\boldsymbol{f}^{(t)}\in\mathbb{R}^{N\times h}}$ (for time step $t$), that sets this weight to a value between 0 and 1 via a sigmoid unit. It is expressed as~\cite[eq. 9.2.1]{zhang2020dive}

\begin{equation}
    {\boldsymbol{f}^{(t)}} = \sigma\!\left(\!{\boldsymbol{x}^{(t)}}{\boldsymbol{W}_{xf}}+\!{\boldsymbol{h}^{(t-1)}}{\boldsymbol{W}_{hf}} \!+\! {\boldsymbol{b}_{f}}\!\right).
    \label{eq003}
\end{equation}
The biases, input weights and recurrent weights for the forget gates are denoted by ${\boldsymbol{b}_{f}}\in\mathbb{R}^{1\times h}$, ${\boldsymbol{W}_{xf}\in\mathbb{R}^{T_{x}\times h}}$, and ${\boldsymbol{W}_{hf}\in\mathbb{R}^{h\times h}}$, respectively.

The external input gate unit ${\boldsymbol{i}^{(t)}\in\mathbb{R}^{N\times h}}$ is computed similar to the forget gate and is expressed as~\cite[eq. 9.2.1]{zhang2020dive}

\begin{equation}
    {\boldsymbol{i}^{(t)}} = \sigma\!\left(\!{\boldsymbol{x}^{(t)}}{\boldsymbol{W}_{xi}}+\!{\boldsymbol{h}^{(t-1)}}{\boldsymbol{W}_{hi}} \!+\! {\boldsymbol{b}_{i}}\!\right),
    \label{eq004}
\end{equation} 
with ${\boldsymbol{W}_{xi}}\in\mathbb{R}^{T_{x}\times h}$ being the input weights, ${\boldsymbol{W}_{hi}}\in\mathbb{R}^{h\times h}$ the recurrent weights, and ${\boldsymbol{b}_{i}}\in\mathbb{R}^{1\times h}$ the bias for the input gate.
The output ${\boldsymbol{h}^{(t)}\in\mathbb{R}^{N\times h}}$ of the LSTM cell, also called hidden state, and the output gate ${\boldsymbol{o}^{(t)}\in\mathbb{R}^{N\times h}}$, are expressed as~\cite[eq. 9.2.4 and 9.2.1]{zhang2020dive}, 
\begin{equation}
    {\boldsymbol{h}^{(t)}} = {\rm{tanh}}\left({\boldsymbol{c}^{(t)}}\right)\odot{\boldsymbol{o}^{(t)}}
    \label{eq005}
\end{equation} 
and
\begin{equation}  
    {\boldsymbol{o}^{(t)}} = \sigma\!\left(\!{\boldsymbol{x}^{(t)}}{\boldsymbol{W}_{xo}}+\!{\boldsymbol{h}^{(t-1)}}{\boldsymbol{W}_{ho}} \!+\! {\boldsymbol{b}_{o}}\!\right),
    \label{eq006}
\end{equation} 
respectively. Again, the input weights, recurrent weights, and bias for the output gate are respectively denoted as ${\boldsymbol{W}_{xo, }}\in\mathbb{R}^{T_{x}\times h}$, ${\boldsymbol{W}_{ho, }}\in\mathbb{R}^{h\times h}$, and ${\boldsymbol{b}_{o}}\in\mathbb{R}^{1\times h}$. The hidden state vector is simply a gated version of the hyperbolic tangent of the memory cell. This ensures that ${\boldsymbol{h}^{(t)}}$ is always between -1 and 1. Whenever the output gate approximates to 1, all information is effectively passed from memory to the predictor, while for the output gate close to 0, all the information is retained within the memory cell and no further processing is performed.

\subsection{Gated Recurrent Unit}
GRUs are a newer generation of RNNs and work similar to LSTMs. Both have a dedicated mechanism composed by gating units to decide when to memorize and when to ignore inputs in the hidden state~\cite{chung2014empirical}. The key difference is that GRUs have only two gates that control the flow of information, namely the reset gate, and update gate. Furthermore, the cell state (memory unit) is not part of its gating unit, and uses only the hidden state ${\boldsymbol{h}^{(t)}\in\mathbb{R}^{N\times h}}$ to transfer information.

\begin{figure}[h]
\centering
\includegraphics[width=3.6in]{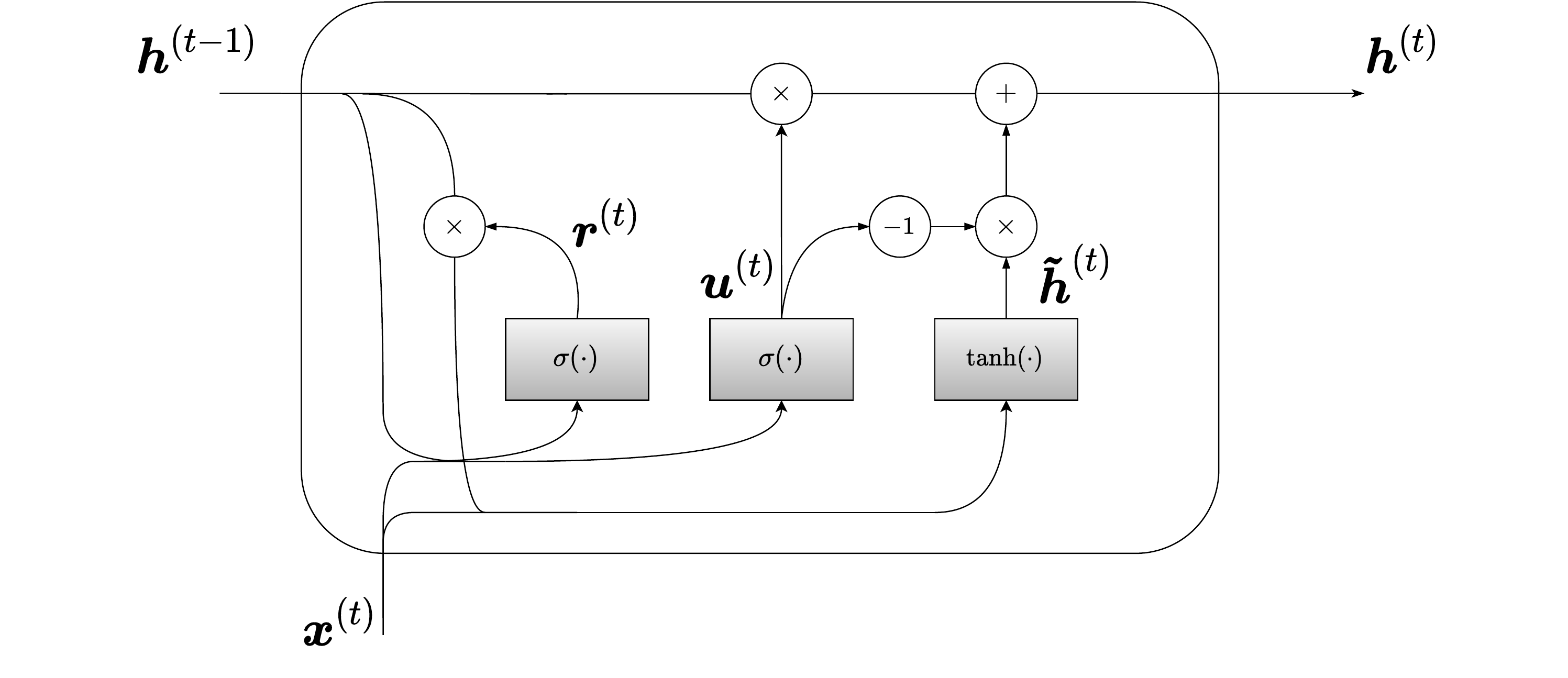}
\caption{Block diagram of the GRU model.}
\label{fig:gru}
\vspace{-0.2cm}
\end{figure}

The core functionality of GRUs rely on a single gating unit simultaneously controlling the forgetting factor and the decision to update the state unit. This update is expressed as~\cite[eq. 10.45]{goodfellow2016deep}
\begin{equation}
    {\boldsymbol{h}^{(t)}} = {\boldsymbol{u}^{(t)}}\odot{\boldsymbol{h}^{(t-1)}}+{(1-\boldsymbol{u}^{(t)})}\odot{\boldsymbol{\tilde{h}}^{(t)}}.
    \label{eq-gru_01}
\end{equation} 
The update gate ${\boldsymbol{u}^{(t)}\in\mathbb{R}^{N\times h}}$ and the reset gate ${\boldsymbol{r}^{(t)}\in\mathbb{R}^{N\times h}}$ are expressed as~\cite[eq. 9.1.1]{zhang2020dive}, 
\begin{equation}
    {\boldsymbol{u}^{(t)}} = \sigma\!\left(\!{\boldsymbol{x}^{(t)}}{\boldsymbol{W}}_{xu}+\!{\boldsymbol{h}^{(t-1)}}{\boldsymbol{W}}_{hu}+\! \boldsymbol{b}^{u}\right)
    \label{eq-gru_02}
\end{equation} 
and
\begin{equation}
    {\boldsymbol{r}^{(t)}} = \sigma\!\left(\!{\boldsymbol{x}^{(t)}}{\boldsymbol{W}}_{xr}+\!{\boldsymbol{h}^{(t-1)}}{\boldsymbol{W}}_{hr}+\! \boldsymbol{b}^{r}\right)
    \label{eq-gru_03}
\end{equation} 
respectively, where ${\boldsymbol{W}_{xu}}, {\boldsymbol{W}_{xr}}\in\mathbb{R}^{T_{x}\times h}$ and ${\boldsymbol{W}_{hu}}, {\boldsymbol{W}_{hr}}\in\mathbb{R}^{h\times h}$ are weight parameters and ${\boldsymbol{b}^{u}}, {\boldsymbol{b}^{r}}\in\mathbb{R}^{1\times h}$ are biases. The current input vector is $\boldsymbol{x^{(t)}}\in\mathbb{R}^{N\times T_{x}}$ with batch size of $N$ and input size $T_{x}$. The current hidden layer containing the GRU outputs is denoted as $\boldsymbol{h^{(t)}}\in\mathbb{R}^{N\times h}$ with $h$ hidden units.

The candidate hidden state $\boldsymbol{\tilde{h}}^{(t)} \in \mathbb{R}^{N\times h}$ at time step $t$ is given as~\cite[eq. 9.1.2]{zhang2020dive}
\begin{equation}
    {\boldsymbol{\tilde{h}}^{(t)}} = {\rm{tanh}}\left({\boldsymbol{x}^{(t)}}{\boldsymbol{W}_{xh}}+{(\boldsymbol{r}^{(t)}\odot\boldsymbol{h}^{(t-1)})\boldsymbol{W}_{hh}}\!+\boldsymbol{b}^{h}\right),
    \label{eq-gru_04}
\end{equation} 
with ${\boldsymbol{W}_{xh}}\in\mathbb{R}^{T_{x}\times h}$, ${\boldsymbol{W}_{hh}}\in\mathbb{R}^{h\times h}$ denoting weight parameters and ${\boldsymbol{b}^{h}}\in\mathbb{R}^{1\times h}$ bias.
\subsection{Convolutional Neural Network}
\begin{figure}[t]
\centering
\includegraphics[width=3.3in]{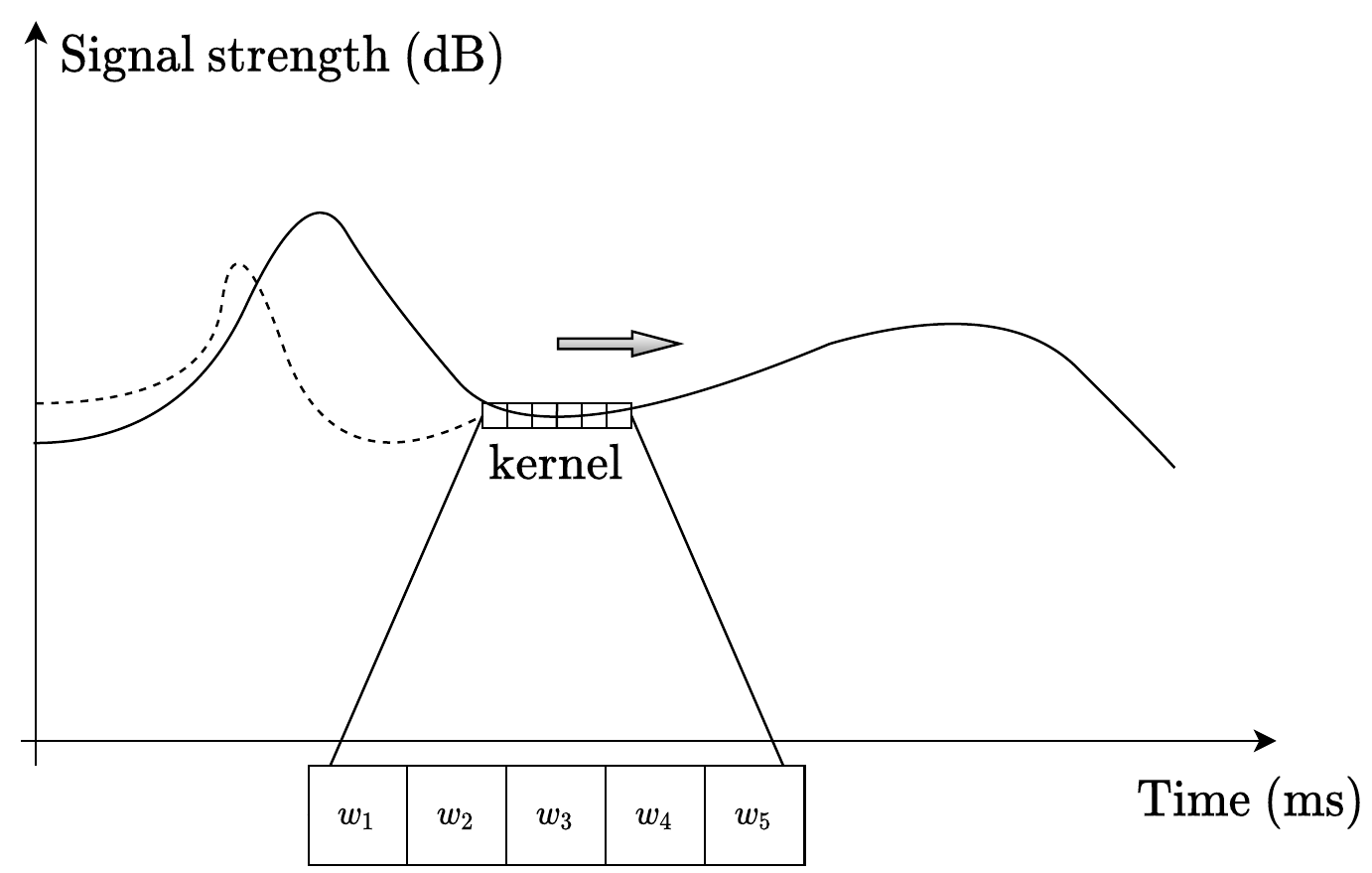}
\caption{Convolution of a time series with a kernel of size 5 producing a representation of the input (dashed line). Here, the solid line indicates the analysed time series and the arrow indicates the direction in which the kernel is being convolved with the signal.}
\label{fig:cnn}
\vspace{-0.2cm}
\end{figure}

The name convolutional neural network indicates that the network employs a mathematical operation called convolution, which is a specialised kind of linear operation. CNNs exploit spatial locality by enforcing a local connectivity pattern between neurons of adjacent layers. As well as this, the convolution of the input with a set of filters (called kernels) is used as the main operation in at least one of its layers. A convolution of a general time series with a kernel of size 5 is shown in~\figref{fig:cnn}. Each kernel convolves with the input producing a representation of the input as an output (this is illustrated as a dashed line in~\figref{fig:cnn}). These representations are then flattened and fed into a feedforward network with one hidden layer with $T_{y}$ hidden units, producing the outputs.

The discrete convolution of an input $x$ with a kernel $w$ results in the output $y$, given by~\cite[eq. 9.3]{goodfellow2016deep}
\begin{equation}
    {y\left(t\right)} = \left(x * w\right)(t) = \sum_{a=-\infty}^{\infty}{x\left(a\right)}{w\left(t-a\right)},
    \label{eq-cnn_01}
\end{equation}
where $*$ represents the convolutional operator, $t$ is the time index, and $x$ and $w$ are defined only on integer $t$.

Unlike GRUs and LSTMs, CNNs are not a type of RNN due to the lack of self-loop mechanisms. Instead, CNNs are well established in the literature and industry as an efficient feature extractor, leading to important progress in computer vision and related tasks~\cite{Krizhevsky2012, he2016, szegedy2015}. 
It is worth highlighting that although CNNs are not RNNs they are known for processing data that have grid-like topology. In particular, one-dimensional convolutional neural networks (1DCNN) are efficient in processing information present in one-dimensional data, such as time series~\cite{goodfellow2016deep}.
The benefit of using 1DCNNs for sequence classification is that they can learn directly from the raw time series data and do not require domain experience to manually design input characteristics. 

\subsection{Implementation Details}
The models used in this paper were built using Ubuntu 20.04.2 LTS and Tensorflow\textsuperscript{\textregistered}\footnote{https://www.tensorflow.org/} 2.5. The data was split into three independent sets, namely training (70\%), validation (20\%) and test sets (10\%). The models were trained and tested in parallel on three computing systems which include:
\begin{enumerate}
    \item a 9th Generation Intel\textsuperscript{\textregistered} Core\textsuperscript{TM} i7-9750H consisting of 6~cores, 16~GB RAM, 512~GB solid state hard drive and an NVIDIA\textsuperscript{\textregistered} GeForce\textsuperscript{\textregistered} GTX 1650 4~GB GDDR5 GPU,
    \item a 10th Generation Intel\textsuperscript{\textregistered} Core\textsuperscript{TM} i9-10885H consisting of 8~cores, 64~GB RAM, 1~TB solid state hard drive and an NVIDIA\textsuperscript{\textregistered} GeForce\textsuperscript{\textregistered} GTX 1650 Ti 4~GB GDDR6 GPU and
    \item a 10th generation Intel\textsuperscript{\textregistered} Core\textsuperscript{TM} i9-10900KF consisting of 10~cores, 64~GB RAM, 2~TB solid state hard drive and an NVIDIA\textsuperscript{\textregistered} GeForce\textsuperscript{\textregistered} RTX 2080~Ti 11~GB GDDR6 GPU.
\end{enumerate}
\begin{table}[t] 
\renewcommand{\arraystretch}{1.2}
\centering
\caption{Parameter space investigated for the deep learning models.}
\label{Table_1a}
\begin{threeparttable}
\resizebox{\columnwidth}{!}{%
\begin{tabular}{|c|c|c|c|c|}
\hline 
 & FFN & 1DCNN & GRU & LSTM\tabularnewline
\hline 
\hline 
Kernels/hidden units & 8 - 256 & 1 - 500 & 1 - 500 & 1 - 500\tabularnewline
\hline 
Layers & 1-2 & 1-2 & 1-2 & 1-2\tabularnewline
\hline 
Input lengths & \multicolumn{4}{c|}{1, 4, 8, 11, 14, 17, 23, 25, 35, 50, 75, 100}\tabularnewline
\hline 
\end{tabular}
}
  \vspace{-0.2cm}
\end{threeparttable}
\end{table}
\begin{table*}[t] 
\renewcommand{\arraystretch}{1.2}
\centering
\caption{Predicted output sequence lengths for D2D measurements in different environments and conditions.}
\label{Table_1}
\begin{threeparttable}
\resizebox{\textwidth}{!}{
\begin{tabular}{|c|l|l|l|l|l|l|l|l|}
\hline
\multirow{2}{*}{\begin{tabular}[c]{@{}c@{}}Time Correlation \\ Function\end{tabular}}     & \multicolumn{2}{c|}{Indoor LOS}                                                                                                                             & \multicolumn{2}{c|}{Indoor NLOS}                                                                                                                            & \multicolumn{2}{c|}{Outdoor LOS}                                                                                                                            & \multicolumn{2}{c|}{Outdoor NLOS}                                                                                                                           \\ \cline{2-9} 
                                                                                          & \begin{tabular}[c]{@{}l@{}}Coherence \\ time (s)\end{tabular}               & \begin{tabular}[c]{@{}l@{}}Output sequence \\ length (samples)\end{tabular} & \begin{tabular}[c]{@{}l@{}}Coherence \\ time (s)\end{tabular}               & \begin{tabular}[c]{@{}l@{}}Output sequence \\ length (samples)\end{tabular} & \begin{tabular}[c]{@{}l@{}}Coherence \\ time (s)\end{tabular}               & \begin{tabular}[c]{@{}l@{}}Output sequence \\ length (samples)\end{tabular} & \begin{tabular}[c]{@{}l@{}}Coherence \\ time (s)\end{tabular}               & \begin{tabular}[c]{@{}l@{}}Output sequence \\ length (samples)\end{tabular} \\ \hline
\multicolumn{1}{|l|}{\begin{tabular}[c]{@{}l@{}}0.1\\ 0.3\\ 0.5\\ 0.7\\ 0.9\end{tabular}} & \begin{tabular}[c]{@{}l@{}}0.017\\ 0.014\\ 0.011\\ 0.008\\ 0.004\end{tabular} & \begin{tabular}[c]{@{}l@{}}17\\ 14\\ 11\\ 8\\ 4\end{tabular}                & \begin{tabular}[c]{@{}l@{}}0.018\\ 0.015\\ 0.012\\ 0.008\\ 0.004\end{tabular} & \begin{tabular}[c]{@{}l@{}}18\\ 15\\ 12\\ 8\\ 4\end{tabular}                & \begin{tabular}[c]{@{}l@{}}0.023\\ 0.019\\ 0.014\\ 0.010\\ 0.005\end{tabular} & \begin{tabular}[c]{@{}l@{}}23\\ 19\\ 14\\ 10\\ 5\end{tabular}               & \begin{tabular}[c]{@{}l@{}}0.016\\ 0.014\\ 0.011\\ 0.008\\ 0.004\end{tabular} & \begin{tabular}[c]{@{}l@{}}16\\ 14\\ 11\\ 8\\ 4\end{tabular}                \\ \hline
\end{tabular}}
  \vspace{-0.2cm}
\end{threeparttable}
\end{table*}

Each of the models mentioned in the previous subsections were trained by optimizing a mean square error (MSE) objective function. The Adam~\cite{Adam} optimization algorithm was used as an adaptive learning rate method with a step size of 0.001. Depending on the configuration of the parameters and computer used, training the deep learning models (i.e., a single experiment) took a maximum of 1 hour whilst the testing phase of the model took only a few minutes. Furthermore, due to the high computational requirement of the deep learning models, the parameter space was extensively investigated over a few months before empirically\footnote{It is worth highlighting that determining the optimal parameter values theoretically for a particular data set is still an open research question. Hence, for the work carried out here, the parameters were determined empirically~\cite{8884240}.} deciding upon the optimal  parameters of the model. Table~\ref{Table_1a} provides the parameter space explored for the deep learning models on which the predictions here are based. 
A batch size of 32 was chosen as no significant improvements were noticed when the batch size was increased from 32 to 64. A dropout size of 0.3 was chosen to reduce overfitting of the model to the training data, and the models were trained for 150 epochs.\footnote{The number of epochs is the number of complete passes through the training data set. Assume that a data set has $x$ number of samples (rows of data), a batch size of $y$ and $z$ epochs. This means that the data set will be divided into $x/y$ batches, each with $y$ samples. The model weights will be updated after each batch of $y$ samples. This also means that one epoch will involve $x/y$ batches or $x/y$ updates to the model. With $z$ epochs, the model will be exposed to or pass through the whole data set $z$ times i.e., a total of $x/y \times z $ batches during the entire training process.}\textsuperscript{,}\footnote{In this study, there was no significant improvement in the performance noticed when the number of epochs was raised above 150.} Besides dropout, early stopping with patience of 15 epochs was used to regularize the models.

As shown in Table~\ref{Table_1}, for each of the D2D data sets the output lengths to be predicted were varied between 4 and 23 samples depending on the coherence time of the channel and time correlation function. According to~\cite{rappaport1996wireless}, the coherence time is defined as the time over which the time correlation function is above 0.5. However, in this study the time correlation function was varied from 0.1 to 0.9 to obtain a range of output lengths for prediction, and to evaluate the prediction performance of the models. For instance, for the indoor LOS measurements when the time correlation function is 0.5, the coherence time of the channel was found to be 11~ms. Since each sample is equal to 1~ms, the corresponding output length is computed to be 11 samples. In the experiments conducted here (at both training and test times), single shot predictions were made where the model predicted out steps time steps in the future, given RSS measurement samples of length between 1 and 100.

\section{Experimental Analysis and Results}
This section discusses the main metrics used for evaluation in this study, the experiments performed on each of the D2D data sets to determine the model and parameters that provide the best prediction performance, and the results.

The main metrics used for evaluation in this study include the mean absolute error (MAE) and root mean squared error (RMSE). MAE measures the average magnitude of the errors in a set of predictions without considering their direction. It is the average over the test sample of the absolute differences between prediction and actual observation where all individual differences have equal weight.
RMSE also measures the average magnitude of the error. It is the square root of the average of squared differences between prediction and actual observation. Since the errors are squared before they are averaged, the RMSE gives a relatively high weight to large errors. This means that the RMSE should be more useful when large errors are particularly undesirable. Both MAE and RMSE express average model prediction error in units of the variable of interest. They are negatively-oriented scores, which means lower values are better. Let $y_{mn}$ be the $m$th test sample for the $n$th prediction step where $n \in \left[1, z\right]$ and $z$ is the total number of prediction steps. Let ${\hat{y}}_{mn}$ be the predicted value of $y_{mn}$. Then, the RMSE and MAE are given by~\eqref{eq007} and~\eqref{eq008} as follows:
\begin{equation}
{\rm{RMSE}}_{n} = \sqrt{\frac{{\sum_{m=1}^{i}\left({\hat{y}}_{mn} - y_{mn}\right)}^2}{i}}
    \label{eq007}
     \end{equation}
     
\begin{equation}
{\rm{MAE}}_{n} = {\frac{{\sum_{m=1}^{i}\left|{\hat{y}}_{mn} - y_{mn}\right|}}{i}}
    \label{eq008}
  \end{equation}
where $i$ is the number of test samples.

\begin{figure*}[t]
\centering
\includegraphics[width=7.12in]{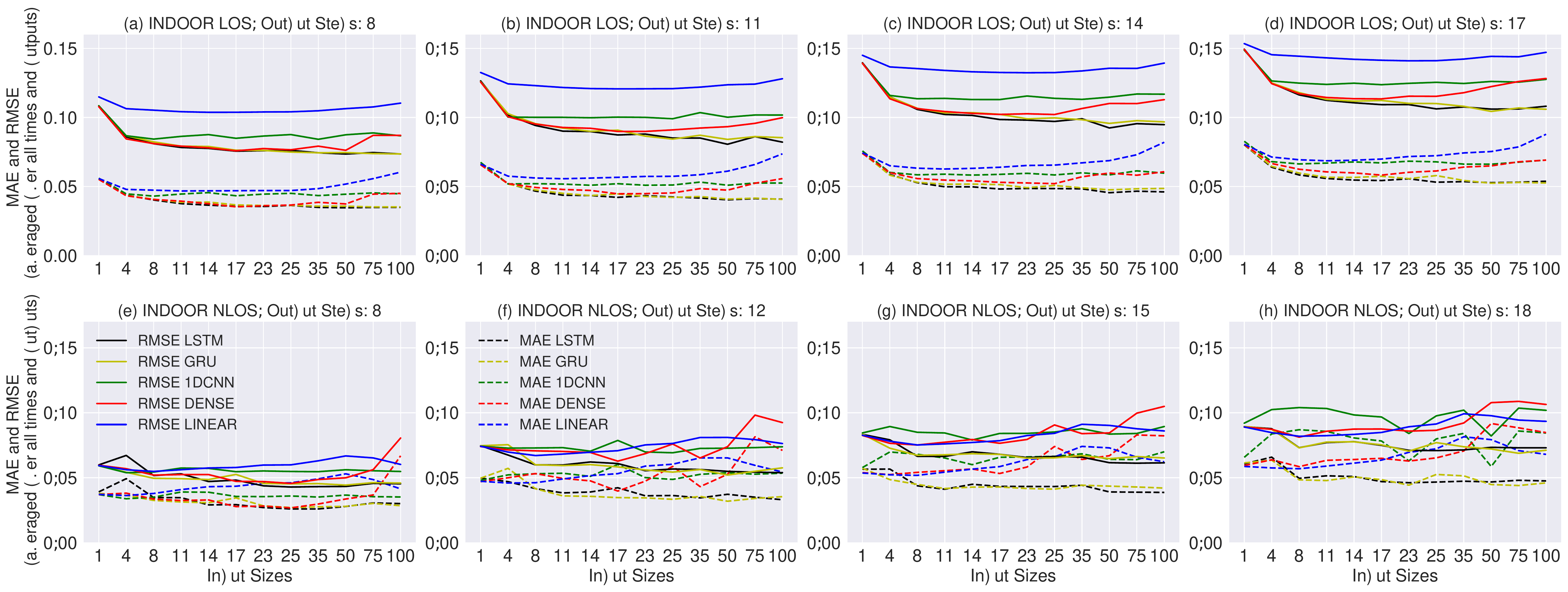}
\caption{Comparison of MAEs and RMSEs of the deep learning models with each other and linear regression for varying input lengths in indoor LOS and NLOS environments. Here, (a), (b), (c) and (d) are obtained for an indoor LOS environment for output lengths 8, 11, 14 and 17, respectively, whilst (e), (f), (g) and (h) are obtained for an indoor NLOS environment for output lengths 8, 12, 15 and 18, respectively. }
\label{fig:img_new1}
\vspace{-0.2cm}
\end{figure*}
\begin{figure*}[t]
\centering
\includegraphics[width=7.12in]{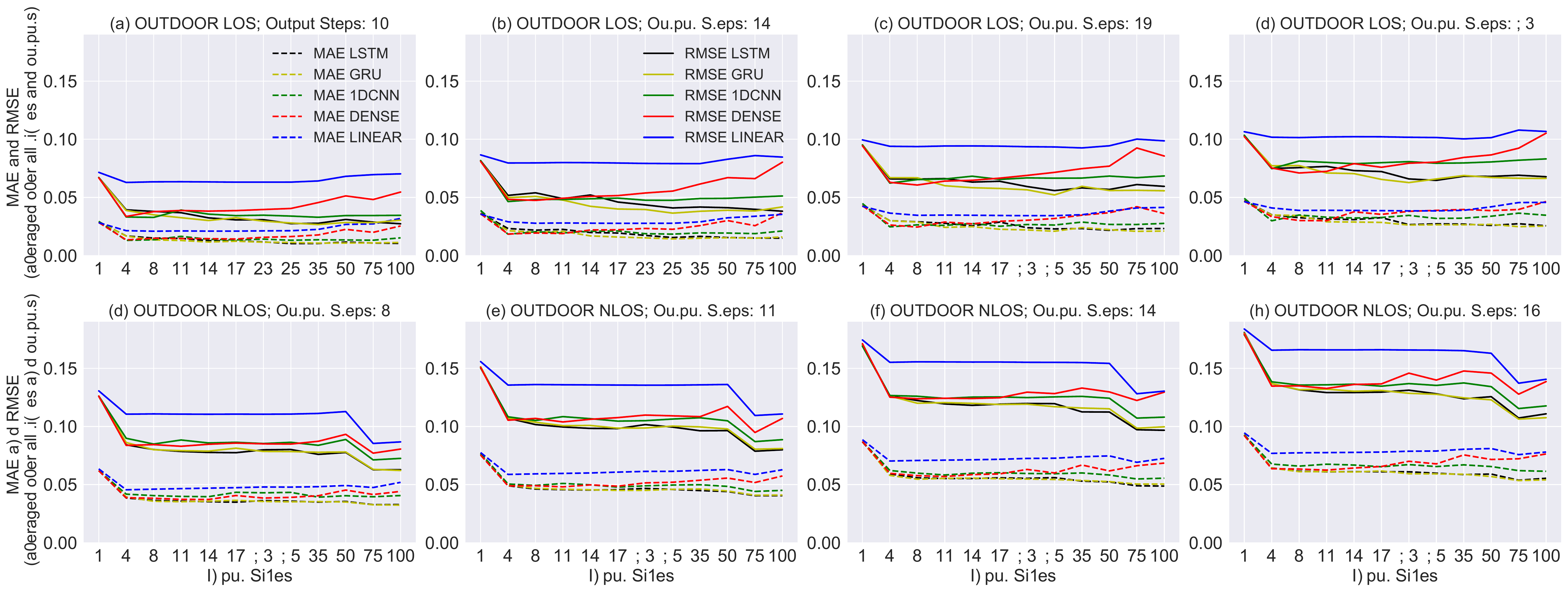}
\caption{Comparison of MAEs and RMSEs of the deep learning models with each other and linear regression for varying input lengths in outdoor LOS and NLOS environments. Here, (a), (b), (c) and (d) are obtained for an outdoor LOS environment for output lengths 10, 14, 19 and 23, respectively, whilst (e), (f), (g) and (h) are obtained for an outdoor NLOS environment for output lengths 8, 11, 14 and 16, respectively.}
\label{fig:img_new2}
\vspace{-0.2cm}
\end{figure*}
\begin{figure*}[t]
\centering
\includegraphics[width=6.6in]{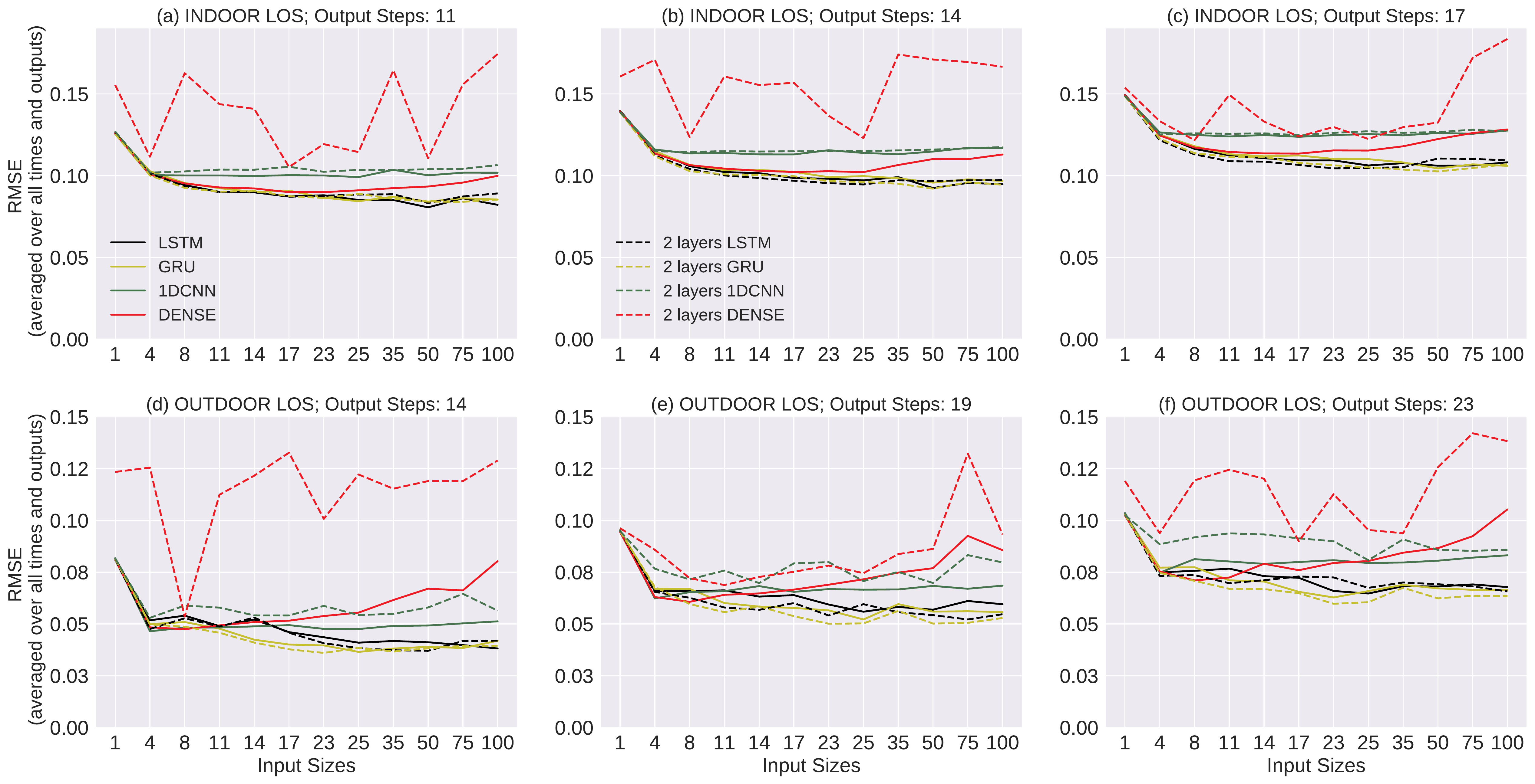}
\caption{Comparing the RMSEs of all deep learning and baseline models, for varying input lengths, for single and multiple LSTM, GRU, 1DCNN and FFN layers in D2D indoor and outdoor LOS environments. Here, (a), (b) and (c) were obtained for an indoor LOS environment for output prediction lengths of 11, 14 and 17 samples, whilst (d), (e) and (f) were obtained for an outdoor LOS environment for output prediction lengths of 14, 19 and 23 samples, respectively.}
\label{fig:img_new3}
\vspace{-0.2cm}
\end{figure*}

Recall that for each data set, the model parameters were tuned by varying the number of stacked layers and hidden units/ kernels in each layer as indicated in Table~I. Following extensive experimentation on the parameter space, in general, it was observed that a single LSTM, GRU and FFN layer with 25 hidden units, and a single 1DCNN layer with 128 kernels and kernel size of 5, provided the best prediction performance across all environments and scenarios. Increasing the number of hidden units to more than 25, and the number of kernels to beyond 128, significantly increased the time taken to train the networks without providing any substantial performance improvements. Likewise, when the number of layers were increased to two, in general, the best prediction performances were obtained when the number of hidden units in each layer for the FFNs were 5; 25 for GRUs and LSTMs; and 64 kernels for 1DCNNs. 

Once the number of hidden units and kernels for the deep learning models were chosen, their MAEs and RMSEs were compared with each other and the linear regression baseline model for varying input and output lengths, for single and multiple layers, across different D2D environments and scenarios. These results are discussed in detail next. 
\begin{figure*}[t]
\centering
\includegraphics[width=6.6in]{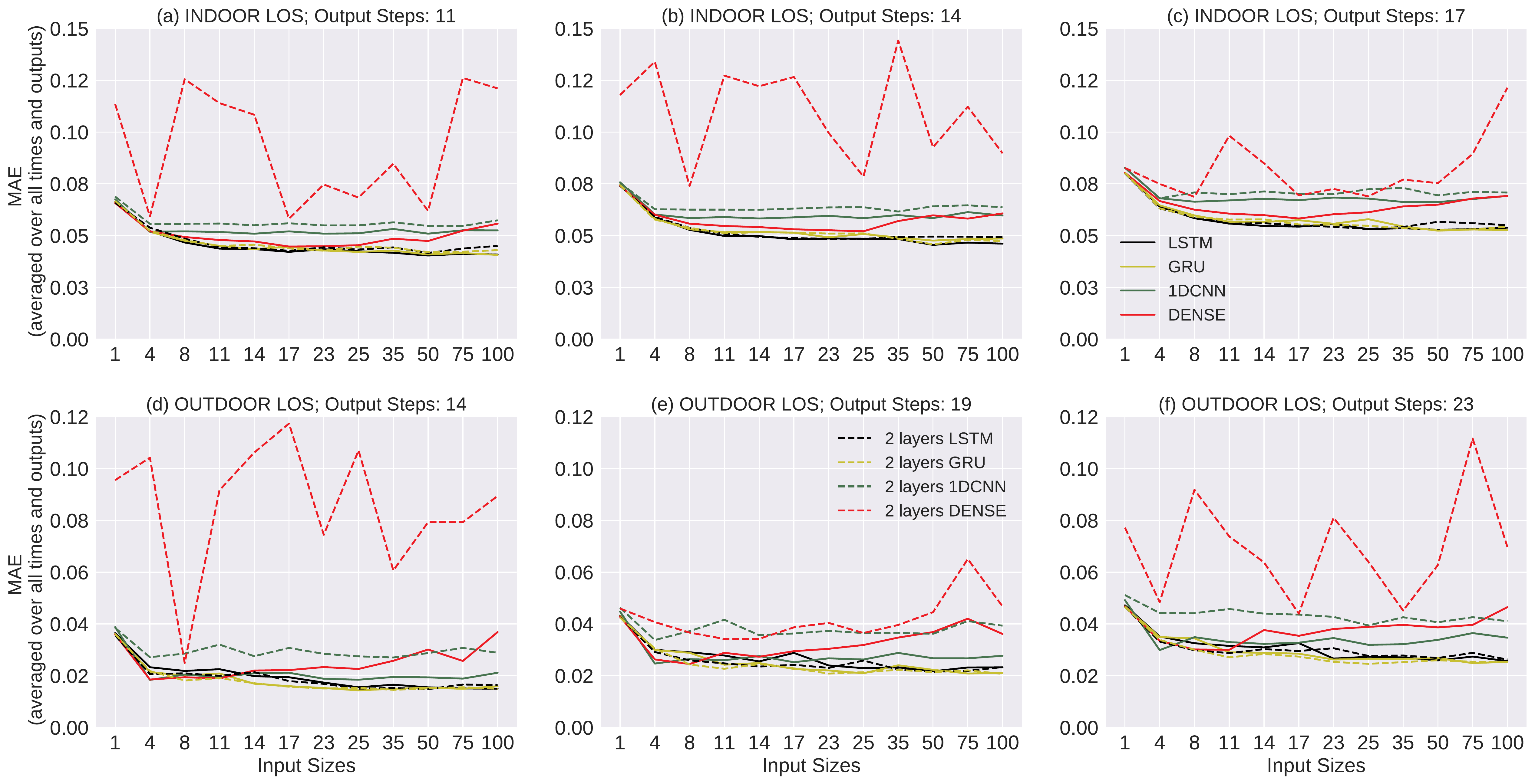}
\caption{Comparing the MAEs of all deep learning and baseline models, for varying input lengths, for single and multiple LSTM, GRU, 1DCNN and FFN layers in D2D indoor and outdoor LOS environments. Here, (a), (b) and (c) were obtained for an indoor LOS environment to predict output lengths of 11, 14 and 17 samples, whilst (d), (e) and (f) were obtained for an outdoor LOS environment to predict output lengths of 14, 19 and 23 samples, respectively.}
\label{fig:img_new4}
\vspace{-0.2cm}
\end{figure*}

\subsection{Comparing Errors vs Input Lengths}

\begin{figure*}[h]
\centering
\includegraphics[width=7.13in]{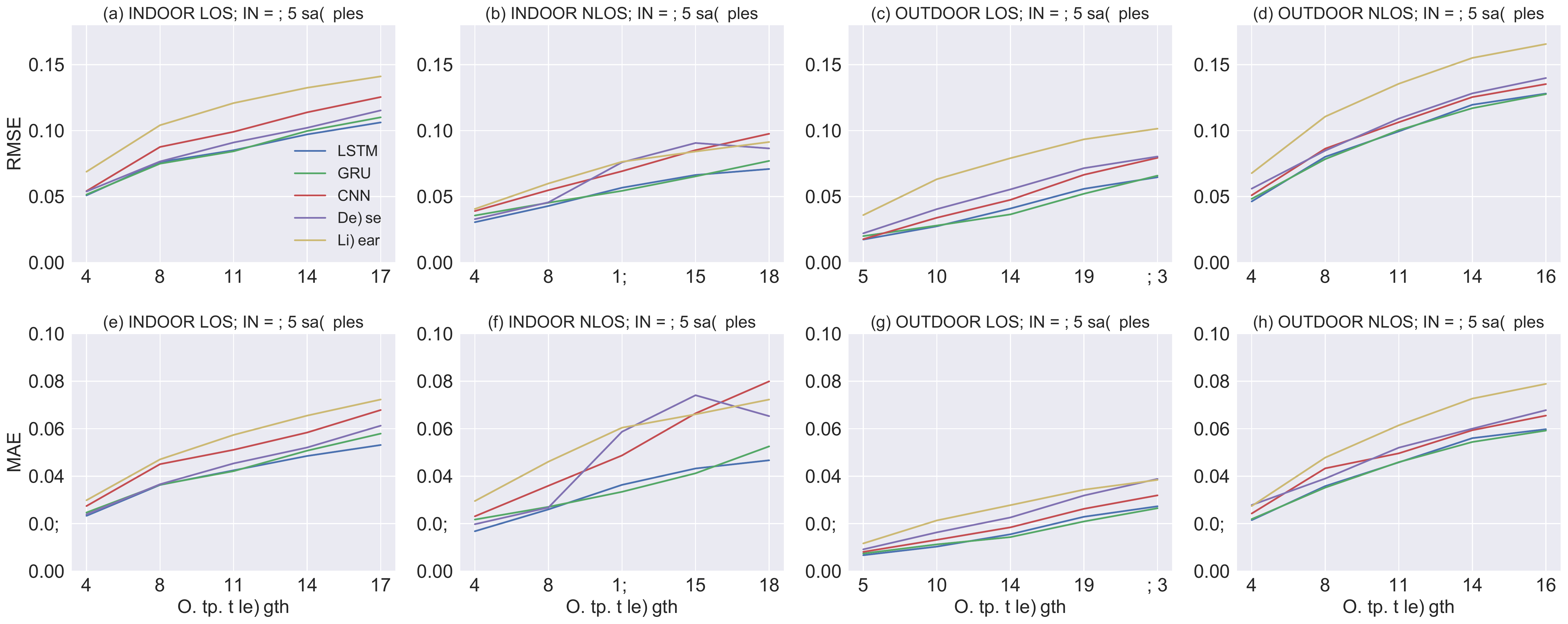}
\caption{Comparing the prediction errors for all deep learning and baseline models, for the optimal number of hidden units/kernels, for different output time steps across all D2D environments and scenarios. The plots have been obtained for a single LSTM, GRU, 1DCNN and FFN layer when the input length was 25 samples. RMSE results are shown for (a) indoor LOS, (b) indoor NLOS, (c) outdoor LOS and (d) outdoor NLOS. MAE results are presented for (e) indoor LOS, (f) indoor NLOS, (g) outdoor LOS, and (h) outdoor NLOS.}
\label{fig:img10}
\vspace{-0.2cm}
\end{figure*}

Figs.~\ref{fig:img_new1} and~\ref{fig:img_new2} compare the RMSEs and MAEs of the deep learning models with each other, and linear regression\footnote{Linear regression is a statistical model that fits the best line to the input data. Similar to the deep neural networks, the baseline also considers a history of 1 - 100 samples to predict the required out steps time steps in the future.}, for a range of input lengths. Figs.~\ref{fig:img_new1}~(a), (b), (c) and (d) were obtained for an indoor LOS environment for output lengths 8, 11, 14 and 17, respectively, whilst Figs.~\ref{fig:img_new1}~(d), (e), (f) and (g) were obtained for an indoor NLOS environment for output lengths 8, 12, 15 and 18, respectively. Likewise, Figs.~\ref{fig:img_new2}~(a), (b), (c) and (d) were obtained for an outdoor LOS environment for output lengths 10, 14, 19 and 23, respectively, whilst Figs.~\ref{fig:img_new2}~(d), (e), (f) and (g) were obtained for an outdoor NLOS environment for output lengths 8, 11, 14 and 16, respectively.

Interestingly, it is observed that a fine-tuned GRU performs very similar to the LSTM. GRUs control the flow of information in essentially the same way as LSTMs. The difference is that LSTMs use a specifically designed memory cell to capture the long-term dependencies in sequences whilst the GRUs use the update gate. Furthermore, these models outperformed FFNs and 1DCNNs for all measured data sets in this study. This points to the importance of accounting for long-term temporal dependencies for channel prediction, which FFNs and 1DCNNs are unable to capture. 
It is also seen that GRUs and LSTMs significantly outperform linear regression in all environments and scenarios. The basic idea behind linear regression is to provide a model which can observe linear trends in the data. It is possible that this baseline model did not perform well here because: 1)~the D2D data sets in this paper are composed of real-world measurements, possibly with nonlinearities introduced due to factors such as the presence of obstacles in the environment, and the direction in which the receiver moved as a result of the random movement undertaken; 2) it very closely follows the trend captured by the previous value, which predicted values may not necessarily follow.


\subsection{Recommendations on Input Length for Coherence Time Prediction}

From~Figs.~\ref{fig:img_new1} and~\ref{fig:img_new2} it is possible to see how little input a model requires to achieve a target prediction performance. For instance, to predict 17 and 23 samples into the future, corresponding to coherence times of 17 and 23~ms in indoor and outdoor LOS environments, respectively, an input length of 25 samples is recommended. Likewise, to predict 17 and 18 samples into the future, corresponding to coherence times of 17~and 18~ms in indoor and outdoor NLOS environments, respectively, input lengths of 25 and 75~samples are recommended. Thus, through these figures, the interested reader can obtain information on the minimum input length required to achieve a target prediction performance for their chosen coherence time given the environments being considered are similar to the ones presented in this work. 
It is also worth highlighting that in most cases, a short input length of around 25 samples was found to achieve similar prediction performance when compared to larger input lengths of 100 samples. Thus, indicating that large input lengths (i.e., knowledge of a large number of past values) may not be always be beneficial. This is intuitive because samples further in the past than the coherence time of the channel are uncorrelated and therefore are less likely to carry as much useful information. 

\subsection{Comparing Errors vs Input Lengths for Multiple Layers' Case}
Observations discussed in the above two subsections for the single layers case also hold when the number of stacked layers is increased to two as demonstrated through~Figs.~\ref{fig:img_new3} and~\ref{fig:img_new4}. Fig.~\ref{fig:img_new3} shows the RMSEs of the deep learning and baseline models vs varying input lengths for single and multiple layers case, whilst Fig.~\ref{fig:img_new4} shows the MAEs of the deep learning and baseline models vs varying input lengths for single and multiple layers case. Furthermore, it can be seen that, just a single LSTM or GRU layer provides good prediction performance and increasing the number of stacked layers will increase the training times without providing any considerable performance benefits.

\subsection{Comparing Errors vs Output Lengths}
\figref{fig:img10} compares the prediction errors for all deep learning and baseline models for different output lengths, across all D2D environments and scenarios. These plots have been obtained for a single LSTM, GRU, FFN and 1DCNN layer when the input length was 25 samples, number of hidden units = 25 and number of kernels = 128. 
As before, it can be seen that LSTMs and GRUs perform comparably across all environments and scenarios. Furthermore, these models significantly outperform FFNs, 1DCNNs and linear regression for all of the D2D data sets considered here. It is also seen that the prediction errors associated with the outdoor LOS scenario were the lowest. This could be because of: 1) the overall low fluctuations in the small scale fading data observed here when compared to the indoor LOS, NLOS and outdoor NLOS cases (see~\figref{fig:img2}), which means that the model has less difficulty making predictions, and 2) the fades observed here are not as deep when compared to the indoor LOS, NLOS and outdoor NLOS cases, again making it easier for the models to predict. 

\subsection{Qualitative Comparison}
\begin{figure}[t]
\centering
\includegraphics[width=3.8in]{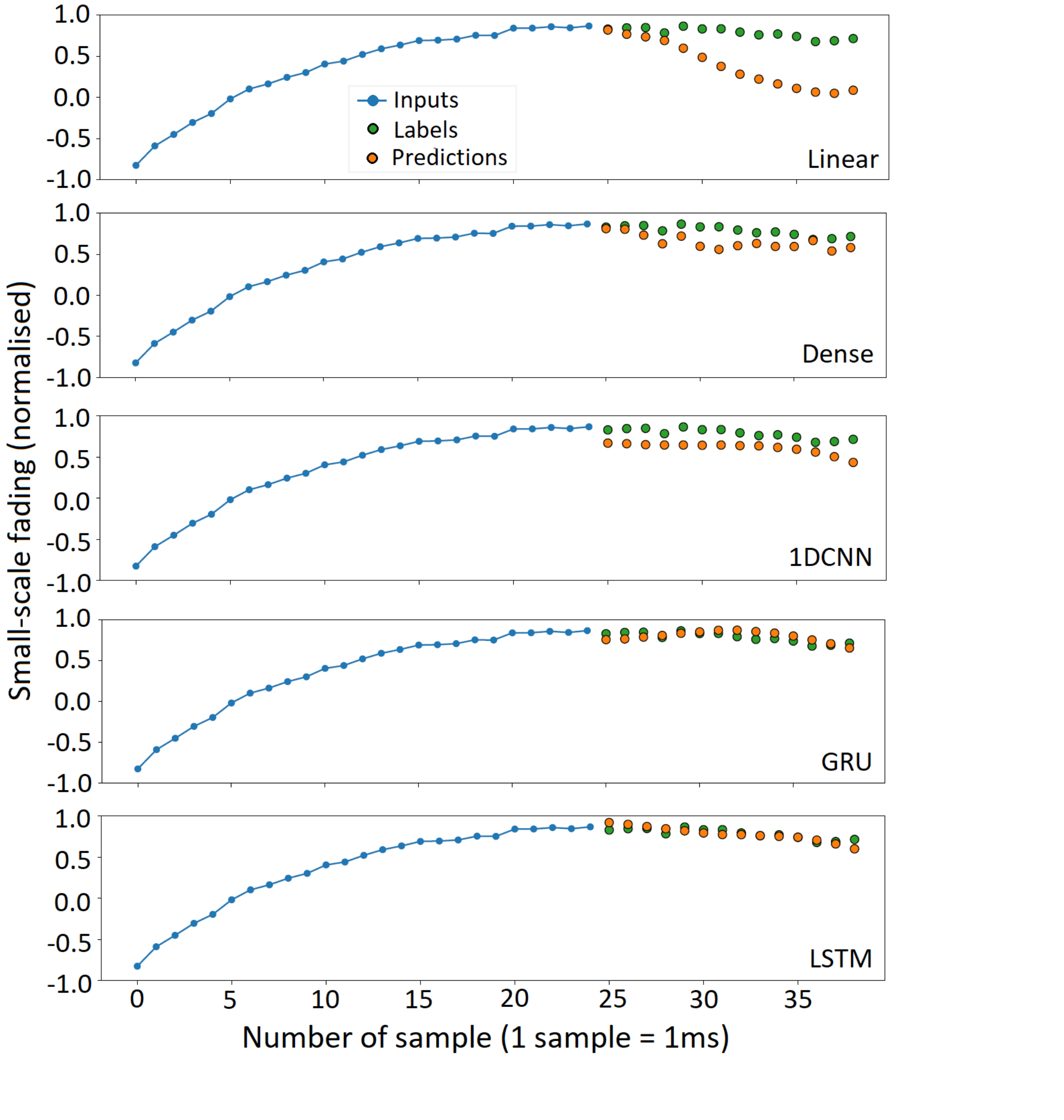}
\caption{Comparison of real and predicted values for outdoor LOS D2D measurements. Here, the output sequence length predicted = 14 samples, input sequence length = 25 samples, number of hidden units = 25 whilst the number of kernels = 128. Min-max normalisation has been applied to the data.}
\label{fig:img_new5}
\vspace{-0.3cm}
\end{figure}

\figref{fig:img_new5} shows a qualitative comparison between the actual and prediction results for linear regression, FFN, 1DCNN, GRU and LSTM models. This figure has been obtained for the outdoor LOS environment and illustrates an input time-frame of 25~ms to predict 14~ms (or 14 samples) in the future. The number of hidden units is equal to 25 whilst the number of kernels is equal to 128. As indicated previously, the linear regression model is only able to capture a low-dimensional slice of the behavior (i.e., it  very  closely  follows  the trend captured by the previous value) resulting in poor prediction performance. GRUs and LSTMs perform the best whilst FFNs and 1DCNNs are the worst performing deep learning models for the given data sets.
\begin{table}[!t] 
\renewcommand{\arraystretch}{1.2}
\centering
\caption{Training times of LSTMs and GRUs for the indoor los and outdoor nlos D2D measurements when the input length is 25 samples.}
\label{Table_7}
\begin{threeparttable}
\begin{tabular}{|c||c|c||c|c|}
\hline 
Environment  & \multicolumn{2}{c||}{Indoor LOS} & \multicolumn{2}{c|}{Outdoor NLOS}\tabularnewline
\hline 
Model & LSTM  & GRU  & LSTM  & GRU \tabularnewline
\hline 
Output (samples) & \multicolumn{2}{c||}{17} & \multicolumn{2}{c|}{16}\tabularnewline
\hline 
Training time (sec) & 121 & 154 & 104 & 90\tabularnewline
\hline 
\hline 
Output (samples) & \multicolumn{2}{c||}{14} & \multicolumn{2}{c|}{14}\tabularnewline
\hline 
Training time (sec) & 121 & 153 & 123 & 101\tabularnewline
\hline 
\hline 
Output (samples) & \multicolumn{2}{c||}{11} & \multicolumn{2}{c|}{11}\tabularnewline
\hline 
Training time (sec) & 121 & 151 & 124 & 112\tabularnewline
\hline 
\hline 
Output (samples) & \multicolumn{2}{c||}{8} & \multicolumn{2}{c|}{8}\tabularnewline
\hline 
Training time (sec) & 134 & 149 & 90 & 115\tabularnewline
\hline 
\hline 
Output (samples) & \multicolumn{2}{c||}{4} & \multicolumn{2}{c|}{4}\tabularnewline
\hline 
Training time (sec) & 114 & 153 & 111 & 66\tabularnewline
\hline 
\end{tabular}
\end{threeparttable}
\end{table}

\subsection{Time Profiling}
Profiling is a way to measure how the models behave in relation to the resources (time and/or memory) they use. It is well known that deep learning models are typically computationally expensive. Thus, quantifying the resource consumption of these models can resolve performance bottlenecks and, ultimately, make them execute faster.
In this subsection, we implement time profiling by comparing the training times of the two best performing deep learning models here, namely LSTM and GRU.  TensorBoard\textsuperscript{\textregistered}~\footnote{https://www.tensorflow.org/tensorboard}, a visualization toolkit of TensorFlow\textsuperscript{\textregistered} was used to profile and track the performance of the models on the device. The device used to evaluate the training times is an NVIDIA\textsuperscript{\textregistered} GeForce\textsuperscript{\textregistered} GTX 1650 4~GB GDDR5 GPU. 

Table~\ref{Table_7} provides the training times of LSTMs and GRUs when the number of hidden units is 25 in the indoor LOS and outdoor NLOS environments. Here, the input length was equal to 25 samples. It is interesting to note that no clear winner was found between LSTMs and GRUs with respect to their training times. For instance, considering the indoor LOS environment and an output length of 17 samples, it can be seen from Table~\ref{Table_7} that the training time associated with the LSTMs was found to be 121~s whilst the GRUs was found to be 154~s. This means that for these parameters in the indoor LOS scenario, the LSTMs trained 27\% faster when compared to the GRUs. Now, considering the outdoor NLOS environment and an output length of 14 samples, it can be seen from Table~\ref{Table_7} that the training time associated with the LSTMs was found to be 123~s whilst the GRUs was found to be 101~s. This means that for these parameters in the outdoor NLOS scenario, the GRUs trained 22\% faster when compared to the LSTMs. Thus, by investigating the prediction performance and training times, it was found that for the D2D measurements considered in this paper, both the GRUs and LSTMs were the best performing models.

\section{Conclusions}
This paper investigated the capabilities of AI-based deep learning models (LSTM, GRU, FFN and 1DCNN) to predict received signal strength variations in D2D communications channels. A thorough investigation was performed on the efficacy of the models to predict different output lengths chosen depending on the coherence time of the channel and time correlation function. It was found that, in general, GRUs and LSTMs  consisting of a single layer with 25 hidden units provided the best prediction performance. Training times of the models were also compared with each other to pick the most suitable model for the D2D data sets considered here. Interestingly, there was no clear winner found between the LSTMs and GRUs. 

The paper also investigated the minimum input length a model requires to achieve a target prediction performance. It was found that to predict 17~and~23~ms into the  future, corresponding to the coherence times observed in indoor and outdoor LOS environments, respectively, an input length of 25~ms was recommended. Likewise, to predict 17 and 18 samples into the future, corresponding to coherence times of 17~and 18~ms in indoor and outdoor NLOS environments,  respectively, input lengths of 25 and 75~samples were recommended.   
This indicates that large input lengths may not always be necessary as samples further in the past than the coherence time of the channel are uncorrelated and  therefore are less likely to carry as much useful information.

\bibliographystyle{IEEEtran}
\bibliography{IEEEabrv,ref}

\begin{IEEEbiography}[{\includegraphics[width=1in,height=1.25 in,clip,keepaspectratio]{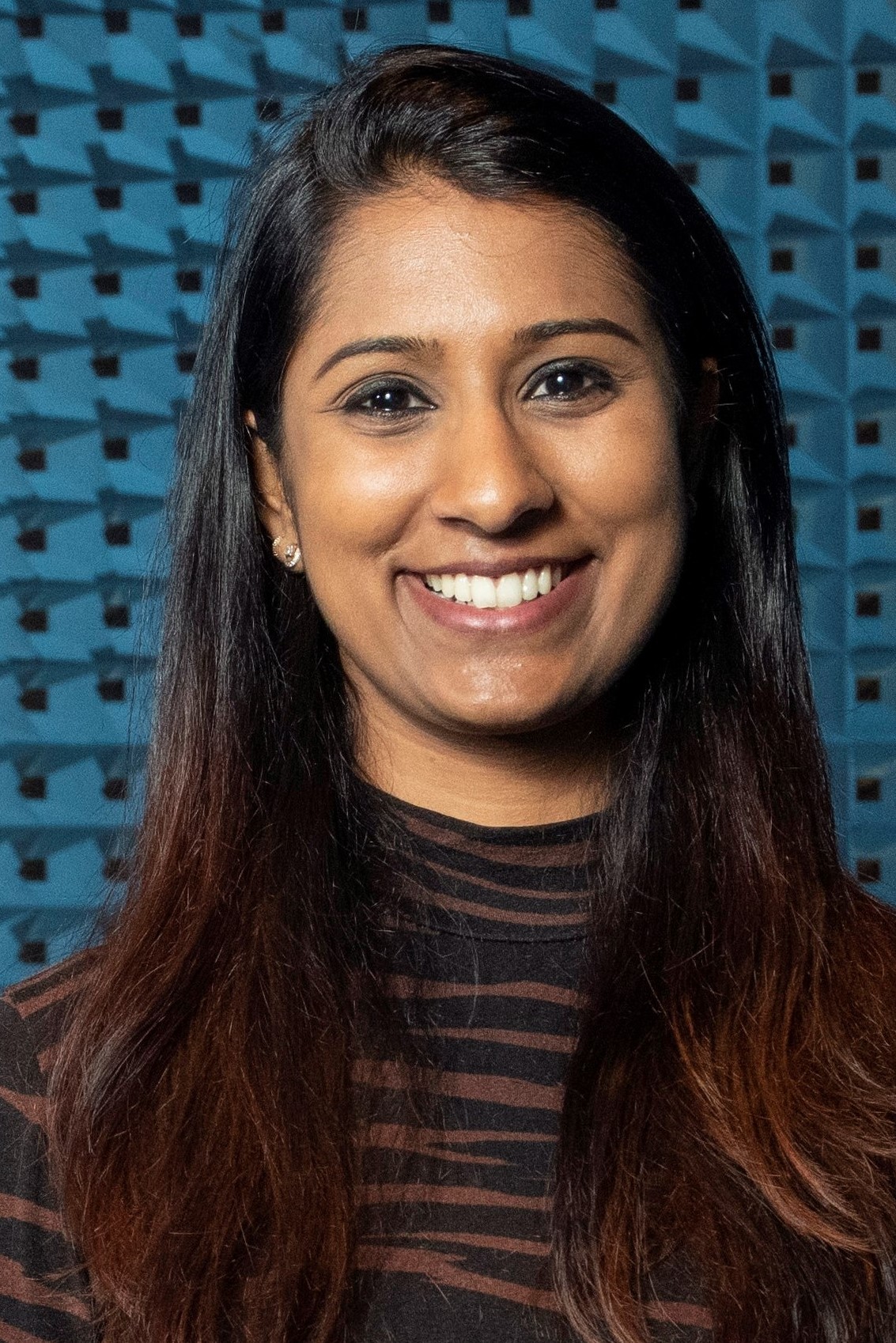}}]{Nidhi Simmons} (S'14--M'18) received the B.E. degree (Hons.) in telecommunications engineering from Visvesvaraya Technological University, Belgaum, India, in 2011, the M.Sc. degree (Hons.) in wireless communications and signal processing from the University of Bristol, U.K., in 2012, and the Ph.D. degree in electrical engineering from Queen's University of Belfast, U.K., in 2018. From 2019 - 2020, she worked as a Research Fellow at Queen's University of Belfast, U.K. She then received a 5-year fellowship from the Royal Academy of Engineering, and currently works as a Royal Academy of Engineering Research Fellow at the Centre for Wireless Innovation, Queen's University Belfast. Her research interests include ultra-reliable low-latency communications, channel characterization and modeling, machine learning for wireless communications and blockchain technology. 
\end{IEEEbiography}

\begin{IEEEbiography}[{\includegraphics[width=1in,height=1.25in,clip,keepaspectratio]{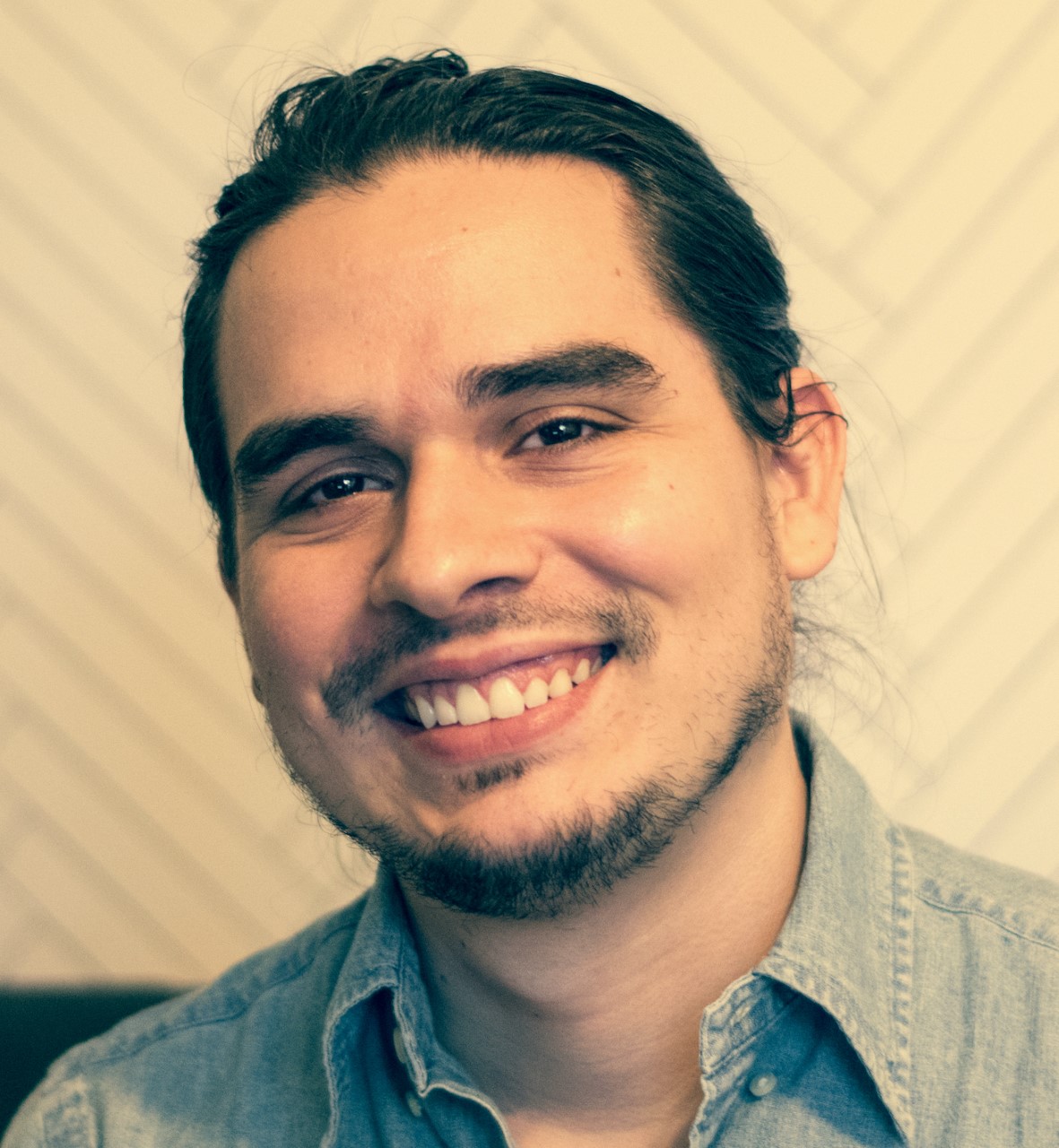}}]{Samuel Borges Ferreira Gomes} received the B.Sc. degree in computer engineering from the Federal University of Ceará (UFC), Brazil, in 2018, and M.Sc. degree in electrical engineering from the State University of Campinas (UNICAMP), Brazil, in 2021, where he is currently working toward the Ph.D. degree. Since 2019, he has been a researcher assistant with the Wireless Technology Laboratory (WissTek), UNICAMP. His research interests include artificial intelligence and wireless communications.
\end{IEEEbiography}

\begin{IEEEbiography}[{\includegraphics[width=1in,height=1.25in,clip,keepaspectratio]{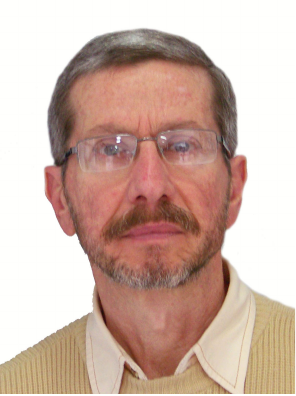}}]{Michel Daoud Yacoub} was born in Brazil, in 1955. He received the B.S.E.E. and M.Sc. degrees from the School of Electrical and Computer Engineering, State University of Campinas, UNICAMP, Campinas, SP, Brazil, in 1978 and 1983, respectively, and the Ph.D. degree from the University of Essex, U.K., in 1988. From 1978 to 1985, he was a Research Specialist in the development of the Tropico digital exchange family with the Research and Development Center of Telebras, Brazil. In 1989, he joined the School of Electrical and Computer Engineering, UNICAMP, where he is currently a Full Professor. He consults for several operating companies and industries in the wireless communications area. He is the author of Foundations of Mobile Radio Engineering (Boca Raton, FL: CRC, 1993), Wireless Technology: Protocols, Standards, and Techniques (Boca Raton, FL: CRC, 2001), and the coauthor of the Telecommunications: Principles and Trends (Sao Paulo, Brasil: Erica, 1997, in Portuguese). He holds two patents. His general research interest includes
wireless communications. 
\end{IEEEbiography}
\begin{IEEEbiography}[{\includegraphics[width=1in,height=1.25in,clip,keepaspectratio]{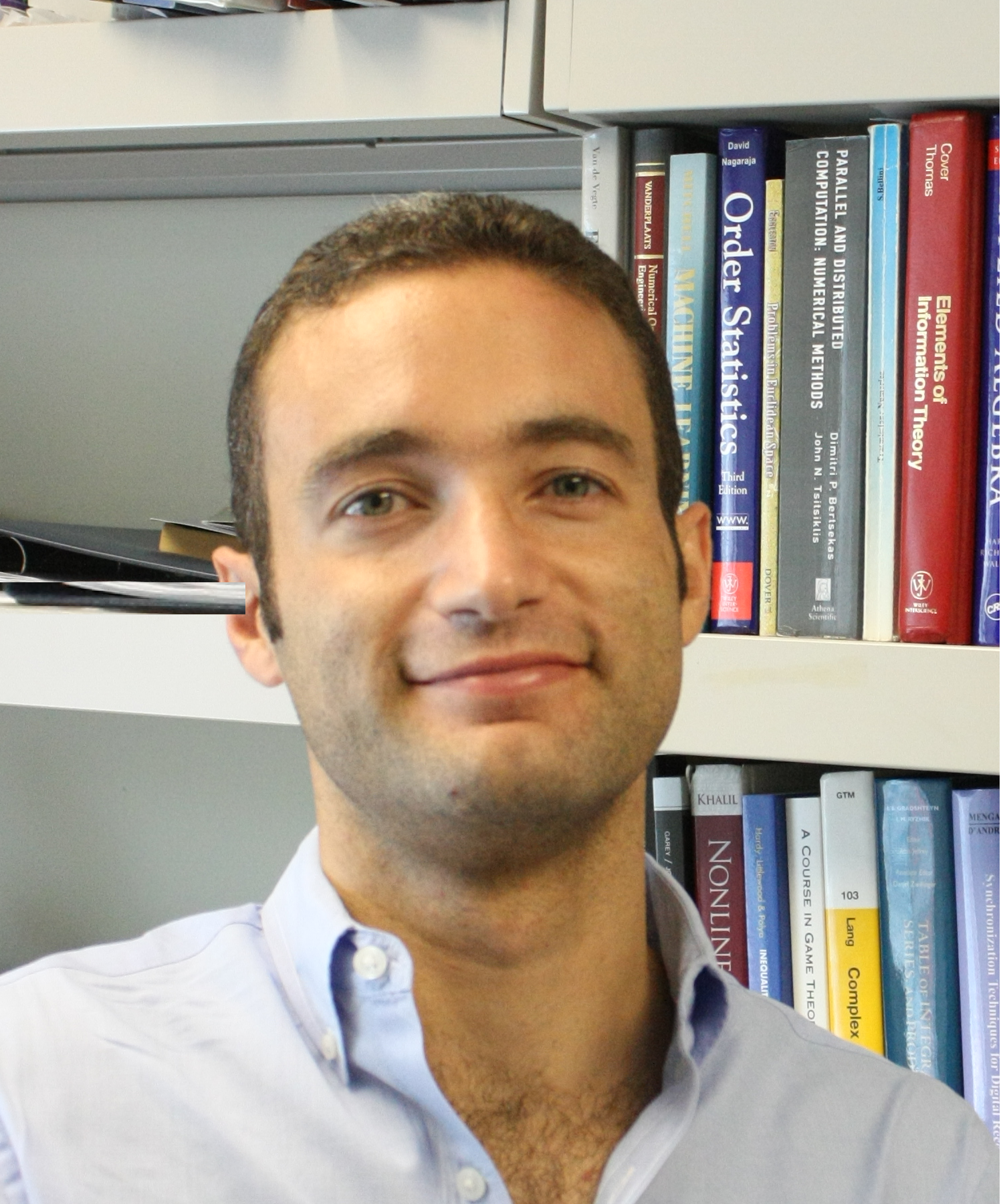}}]{Osvaldo Simeone} (Fellow, IEEE) is a Professor of Information Engineering with the Centre for Telecommunications Research at the Department of Engineering of King's College London, where he directs the King's Communications, Learning and Information Processing lab. He received an M.Sc. degree (with honors) and a Ph.D. degree in information engineering from Politecnico di Milano, Milan, Italy, in 2001 and 2005, respectively. From 2006 to 2017, he was a faculty member of the Electrical and Computer Engineering (ECE) Department at New Jersey Institute of Technology (NJIT), where he was affiliated with the Center for Wireless Information Processing (CWiP). His research interests include information theory, machine learning, wireless communications, and neuromorphic computing. Dr Simeone is a co-recipient of the 2022 IEEE Communications Society Outstanding Paper Award, 2021 IEEE Vehicular Technology Society Jack Neubauer Memorial Award, 2019 IEEE Communication Society Best Tutorial Paper Award, the 2018 IEEE Signal Processing Best Paper Award, the 2017 JCN Best Paper Award, the 2015 IEEE Communication Society Best Tutorial Paper Award and of the Best Paper Awards of IEEE SPAWC 2007 and IEEE WRECOM 2007. He was awarded a Consolidator grant by the European Research Council (ERC) in 2016. His research has been supported by the U.S. NSF, the ERC, the Vienna Science and Technology Fund, as well as by a number of industrial collaborations. He currently serves in the editorial board of the IEEE Signal Processing Magazine and is the chair of the Signal Processing for Communications and Networking Technical Committee of the IEEE Signal Processing Society. He was a Distinguished Lecturer of the IEEE Information Theory Society in 2017 and 2018, and he is currently a Distinguished Lecturer of the IEEE Communications Society. Dr Simeone is a co-author of a textbook (to be published by Cambridge University Press), two monographs, two edited books, and more than 160 research journal papers. He is a Fellow of the IET and of the IEEE. 
\end{IEEEbiography}

\begin{IEEEbiography}[{\includegraphics[width=1in,height=1.25in,clip,keepaspectratio]{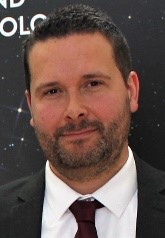}}]{SIMON L. COTTON} (S'04--M'07--SM'14) received the B.Eng. degree in electronics and software from Ulster University, Ulster, U.K., in 2004, and the Ph.D. degree in electrical and electronic engineering from the Queen's University of Belfast, Belfast, U.K., in 2007. He is currently a Professor of Wireless Communications and the Director of the Centre for Wireless Innovation (CWI), Queen's University Belfast. He has authored and co-authored over 140 publications in major IEEE/IET journals and refereed international conferences, two book chapters, and two patents. Among his research interests are cellular device-to-device, vehicular, and body-centric communications. His other research interests include radio channel characterization and modeling, and the simulation of wireless channels. He was a recipient of the H. A. Wheeler Prize, in 2010, from the IEEE Antennas and Propagation Society for the best applications journal paper in the IEEE Transactions on Antennas and Propagation, in 2009, and the Sir George Macfarlane Award from the U.K. Royal Academy of Engineering in recognition of his technical and scientific attainment since graduating from his first degree in engineering, in 2011.
\end{IEEEbiography}

\begin{IEEEbiography}[{\includegraphics[width=1in,height=1.25in,clip,keepaspectratio]{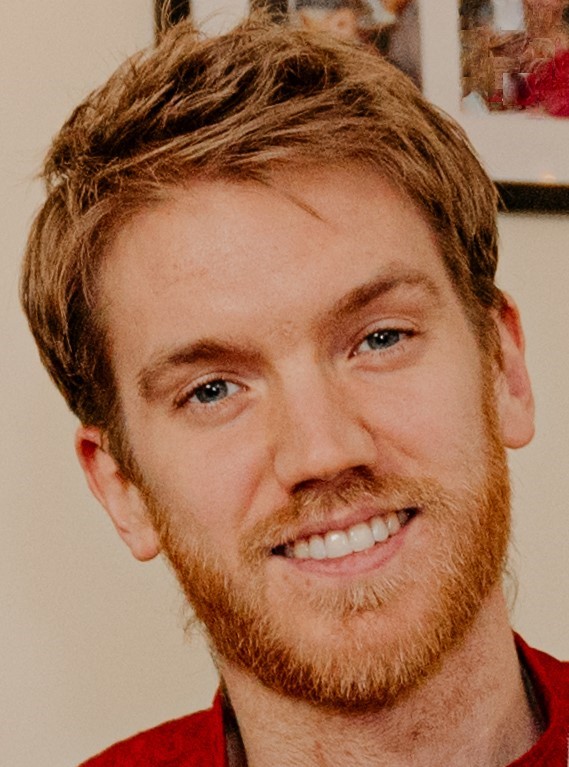}}]{David E. Simmons} received the BSc in mathematics from the University of Central Lancashire, in 2011, the M.Sc. degree in communications engineering from the University of Bristol, U.K., in 2012, and D.Phil. degree in Engineering from the University of Oxford, U.K., in 2016. His research has focused on studying the information theoretic properties of relay networks as they scale. During his D.Phil. studies, he was a recipient of the Best Paper Award at the 23rd edition of EUCNC’14. From 2016 to 2017, he worked as a PDRA within the Networked Quantum Information Technologies group at the University of Oxford. From 2018 until present, he has worked as a Senior AI/ML Research Scientist, Senior Software Engineer, and Engineering Team Lead in two startup companies in Belfast, U.K. His research interests include communication and network theory, information theory, AI/ML, and cryptography.
\end{IEEEbiography}

\EOD

\end{document}